\documentclass[journal]{IEEEtran}
\IEEEoverridecommandlockouts

\usepackage{url}
\usepackage{hyperref}

\usepackage{csquotes}

\ifCLASSINFOpdf
  \usepackage[pdftex]{graphicx}
  \usepackage{caption}
\else
\fi
\usepackage{subcaption}

\usepackage{graphicx}
\usepackage{etoolbox}
\newcommand{\insertfig}{
    \setcounter{figure}{0}
    \includegraphics[width=\textwidth]{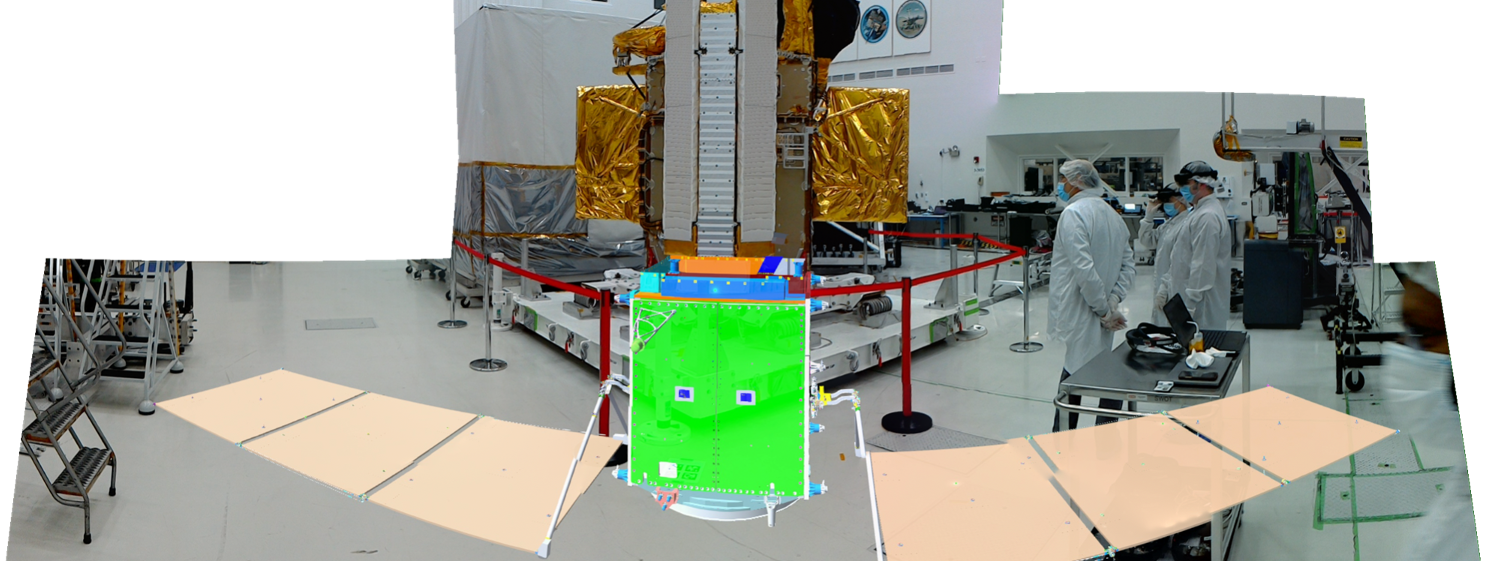}
    \captionof{figure}{ProtoSpace has been used in a variety of use cases, including in JPL’s Spacecraft Assembly Facility as shown here in which a virtual CAD model of the SWOT (Surface Water and Ocean Topography)~\cite{fu2024surface} Spacecraft Bus is seen in augmented reality, aligned directly underneath the physical SWOT Payload Module. In this use case, the Payload Module was physically at JPL and engineers wanted to see how the Spacecraft Bus would align with it during a future assembly. This panoramic photo was stitched together from several mixed-reality captured photos from a HoloLens running ProtoSpace; note that stitching artifacts are created during the panoramic stitching process and are not related to ProtoSpace.}
    \label{fig:swot}
}

\makeatletter
\apptocmd{\@maketitle}{\centering\insertfig}{}{}
\makeatother

\begin{document}
%

\title{A Review of 10 Years of ProtoSpace: \\
Spacecraft CAD Visualization in \\
Collaborative Augmented Reality}

%
%
%

\author{\IEEEauthorblockN{Benjamin Nuernberger\IEEEauthorrefmark{1},
Samuel-Hunter Berndt\IEEEauthorrefmark{2}, 
Robert Tapella\IEEEauthorrefmark{3},\\
Laura Mann\IEEEauthorrefmark{4},
Aaron Plave\IEEEauthorrefmark{5},
Sasha Samochina\IEEEauthorrefmark{6},
Victor X. Luo\IEEEauthorrefmark{7}\\
\IEEEauthorblockA{
Jet Propulsion Laboratory,
California Institute of Technology,  Pasadena, CA USA\\
\IEEEauthorrefmark{1}benjamin.nuernberger@jpl.nasa.gov,
\IEEEauthorrefmark{2}samuel-hunter.berndt@jpl.nasa.gov,
\IEEEauthorrefmark{3}tapella@hotmail.com,\\
\IEEEauthorrefmark{4}laura.mann@jpl.nasa.gov,
\IEEEauthorrefmark{5}aaron.plave@jpl.nasa.gov,
\IEEEauthorrefmark{6}sasha@some-thing.co,
\IEEEauthorrefmark{7}victor@jpl.nasa.gov}}
\thanks{\copyright~2025 California Institute of Technology. Government sponsorship acknowledged.}
\thanks{Each author's contribution to this work was done during their tenure at the Jet Propulsion Laboratory.}
}

\maketitle

\IEEEpubidadjcol


\begin{abstract}
ProtoSpace is a custom JPL-built platform to help scientists and engineers visualize their CAD models collaboratively in augmented reality (AR) and on the web in 3D. In addition to this main use case, ProtoSpace has been used throughout the entire spacecraft mission lifecycle and beyond: ventilator design and assembly; providing AR-based instructions to astronauts in-training; educating the next generation on the process of spacecraft design; etc. ProtoSpace has been used for a decade by NASA missions—including Mars Perseverance~\cite{farley2020mars}, Europa Clipper~\cite{pappalardo2024science}, NISAR~\cite{kellogg2020nasa}, SPHEREx~\cite{crill2020spherex}, CAL~\cite{aveline2020observation}, and Mars Sample Return~\cite{muirhead2020mars}—to reduce cost and risk by helping engineers and scientists fix problems earlier through reducing miscommunication and helping people understand the spatial context of their spacecraft in the appropriate physical context more quickly. This paper will explore how ProtoSpace came to be, define the system architecture and overview—including HoloLens and 3D web clients, the ProtoSpace server, and the CAD model optimizer—and dive into the use cases, spin-offs, and lessons learned that led to 10 years of success at NASA’s Jet Propulsion Laboratory.

\end{abstract}


%
\IEEEpeerreviewmaketitle

\section{Introduction}
%
%
%
%
\IEEEPARstart{D}{esigning}, building, and assembling multi-million dollar spacecraft comes with both inherent challenges and risks as well as enormous potential for once-in-a-lifetime scientific breakthroughs and discovery. 
A key challenge in this process is the difficulty of understanding and communicating spatial information about the spacecraft design and its integration and assembly process. 
Typically, engineers view this inherently 3D information on 2D screens, many times taking screenshots of particular views that they will use to communicate with others. 
The problem is that such 2D projections of the 3D information can introduce an extra layer of cognitive complexity~\cite{dan2017eeg, li2024comparative} possibly increasing the time to understand the spatial information. 
This increases the risk that there will be an inaccurate understanding of a particular situation, or that engineers may miss a problem entirely that should be fixed in the design phase to avoid the large cost of rework later on during the assembly and integration phases. 
Ultimately, such problems can lead to increased labor cost, instability in cost and schedule, and increased risk overall. 

ProtoSpace was built to help alleviate such issues by allowing engineers to view the spacecraft 3D CAD models in stereoscopic augmented reality with relative ease, especially in collaboration scenarios. ProtoSpace is a platform custom built at NASA’s Jet Propulsion Laboratory (JPL) that consists of a backend server (to host CAD model data and facilitate real-time session information) and frontend clients (to view the CAD model). Instead of sitting around a conference room table with everyone looking at their respective computers or staring at a projector screen, engineers can now all stand up and see the full-scale CAD model stereoscopically in augmented reality by using ProtoSpace (see Figure~\ref{fig:clipper-pdr}). Instead of having no way to view a not-yet-built portion of a spacecraft with respect to a built-portion of the spacecraft, now with ProtoSpace cleanroom engineers can see the not-yet-built portion as a virtual CAD model aligned to the physically built-portion in augmented reality (see Figure~\ref{fig:swot}). While ProtoSpace has been showcased previously in various ways over the years~\cite{hanhe2018risk,berndt2023universe,protospaceYouTube,bedrosian2025roman}, this paper provides a fresh take on ProtoSpace and makes the following contributions:
\begin{itemize}
    \item A comprehensive overview of the ProtoSpace platform, including system architecture, user interface design, and use cases that ProtoSpace targets
    \item Details of both design choices and technical optimizations that were key to making ProtoSpace work well for the past 10 years at JPL
    \item Case studies of how ProtoSpace was used successfully by a variety of missions, with quotes from end users
    \item A set of lessons learned and future outlooks for the technology overall
\end{itemize}

The rest of the paper is structured as follows. We first begin by reviewing related work and then describe how ProtoSpace first began. Next, we describe the overall platform as well as the three main use cases of ProtoSpace. Major system details follow and then the case studies. One-off spin-offs of the project come next, followed by lessons learned, and finally the future outlook and conclusion.

 

\section{Related Work}
\subsection{XR CAD Viewing}
\label{sec:related-work-xr-cad-viewing}
The history of viewing computer-aided-drawings (CAD) in extended reality (XR) lines up perfectly with the history of augmented reality more broadly. 
In the 1960’s, Ivan Sutherland’s Sketchpad work~\cite{sutherland1964sketch}, largely considered as the genesis of CAD~\cite{kasik2005ten}, was also a precursor to the first augmented reality head’s up display also produced by Sutherland~\cite{sutherland1968head}. 
Later in 1990, Caudell and Mizell coined the term Augmented Reality~\cite{thomas1992augmented}, in the context of manufacturing for airplanes, even describing how ``packet radio could be used to download CAD/CAM data on the fly'' which is very similar to how ProtoSpace eventually was built by downloading CAD data on the fly over WiFi.

Since then, over the past few decades, there has been a proliferation of systems built for the purpose of viewing CAD in AR. 
Reiners et al.~\cite{reiners1998augmented} demonstrated an augmented reality doorlock assembly tool, even describing how they reduced the model’s complexity so it can be rendered in the display. 
In terms of specific use cases with visualizing CAD, XR has been utilized for Building Information Modeling (BIM)~\cite{wang2014integrating}, engineering design reviews~\cite{wolfartsberger2017virtual}, the Architecture Engineering and Construction (AEC) field~\cite{wen2020using}, and for authoring CAD models themselves~\cite{feeman2018exploration}.
We refer the reader to recent surveys that cover the topic of XR related to industry 4.0 and manufacturing, both of which are highly relevant to visualizing CAD models in XR~\cite{de2020survey, eswaran2022challenges, cardenas2022extended, fang2023head}.

Finally, we make note of a few inherent challenges for viewing CAD models in XR, which have been previously noted by others~\cite{raposo2006towards}: 
\begin{itemize}
    \item The loss of semantic metadata information when converting from native CAD formats (e.g., STEP files) to formats conducive for visualizing in an XR headset (e.g., FBX). In ProtoSpace, we attempt to preserve the piece-part (node) metadata, especially the names and types of nodes.
    \item One-way conversion of CAD data not allowing for review feedback and edits to easily work their way back into the original CAD system. ProtoSpace did not address this issue.
    \item Massive CAD models being made up of tens or hundreds of millions of polygons, which are typically one or more orders of magnitude too many for high frame rate rendering on untethered XR headsets. ProtoSpace addressed this issue through user-interactive model reduction and various real-time rendering optimizations; see Sections~\ref{sec:web-server-backend} and~\ref{sec:rendering}, respectively. Such approaches have been taken by others as well~\cite{lorenz2016cad,unity}. Another popular approach in recent years is offloading the rendering to the cloud (remote rendering).
\end{itemize}

\begin{figure*}[th]
    \centering
    \includegraphics[width=0.75\textwidth]{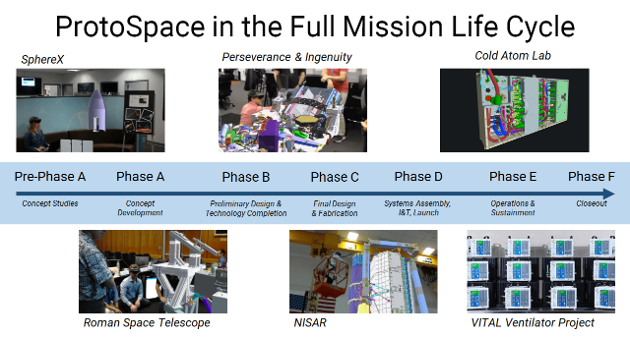}
    \caption{ProtoSpace has been used throughout the full mission lifecycle at JPL.}
    \label{fig:full-lifecycle}
\end{figure*}

\subsection{XR for Space Applications}
NASA has used XR in a variety of ways over the years~\cite{zhivotovskiy2024reality}.
The NASA Ames Research Center (ARC) created one of the first early virtual reality (VR) headsets called the Virtual Interface Environment Workstation (VIEW)~\cite{fisher1988virtual, fisher2016nasa}.
VR has been used with astronauts for training and maintenance onboard the ISS ~\cite{garcia2020training,nuernberger2020under,kellogg2023augmented,paddock2023nasa,weiss2024augmented}. 
XR has also been used for earth science visualization~\cite{2018AGUFMIN51D0608R, grubb2024development}, controlling rovers~\cite{frantz2025xrov}, spacecraft magnetic field visualization~\cite{nuernberger2023visualizing}, and immersive viewing of the surface Mars~\cite{abercrombie2017onsight}.
In recent years, some have also investigated using ``metaverse'' technologies for enhancing overall work~\cite{berndt2023universe}.
While ProtoSpace has focused mostly on the design and building of spacecraft hardware, there have been a few instances where it was used for other use cases, such as cubesat mission design (Section~\ref{sec:teamxc}) and immersive procedure assistance for astronauts (Section~\ref{sec:procedures}).


\section{The Genesis of ProtoSpace}
In 2014, we began co-development of the first AR applications on the Microsoft Hololens platform—starting with OnSight (enable NASA scientists to walk virtually on Mars)~\cite{onsightVideo,abercrombie2017onsight} and Sidekick (real-time expert assistance from ground control to ISS)~\cite{sidekickWiki}.  ProtoSpace was conceived in late 2015—allowing spacecraft engineers to collaboratively design in virtual 3D space.  Primary development focused on CAD model to mobile-device ready model conversion, basic gestural control in 3D space, and scene synchronization. Eventually, more complex features were developed to allow spatial procedures to be aligned and overlaid on physical objects—providing a self-service complement to (the more real-time) Sidekick.

There were initially three general categories of use cases: (1) Mechanical engineers designing spacecraft; (2) Formulation and proposal showcase; (3) Spatial procedure overlay. Since then, ProtoSpace has focused mostly on the first two of these (see Section~\ref{sec:three-main-use-cases}) while also having done prototyped and tested procedure assistance in a couple of scenarios (see Section~\ref{sec:procedures}).

\section{Summary Overview}
In this section, we give a brief overview of the platform, describe how ProtoSpace has been used and added value through the entire mission lifecycle, and cover the three main use cases of ProtoSpace over the past 10 years.

\subsection{Platform Overview}
Using ProtoSpace consists of first uploading a CAD model to the ProtoSpace website.
The backend server ingests the CAD models and makes it ready for interactive model reduction.
Through a simple user interface, the end user can selectively choose which parts of the model to keep at full fidelity while reducing the complexity of other parts.
Once the model is ready, users can either start a new session through the website or through the ProtoSpace HoloLens app.
The main augmented reality experience is achieved through the HoloLens client.
Details of the entire system are described in Section~\ref{sec:details}.

\subsection{Throughout the Mission Lifecycle}
As shown in the Figure~\ref{fig:full-lifecycle}, ProtoSpace has been utilized throughout the entire NASA mission lifecycle. From Pre-Phase A (e.g., see Section~\ref{sec:spherex}) all the way through Phase E in operations (e.g., see Sections~\ref{sec:vital} and~\ref{sec:cal-gifs}), ProtoSpace has been used before hardware was fully designed, through the final designs and into assembly and test, and even into operations.

\subsection{Three Main Use Cases}
\label{sec:three-main-use-cases}
The following three main use cases are considered production ready use cases that have been utilized numerous times at JPL. In many cases, end users of the platform have utilized ProtoSpace entirely by themselves. On the other hand, ProtoSpace has also been used in a variety of one-off use cases that were quite successful but were not considered part of the main product; see Section~\ref{sec:one-offs}.

\subsubsection{Designing Spacecraft}
\begin{figure}[h]
    \centering
    \includegraphics[width=1\linewidth]{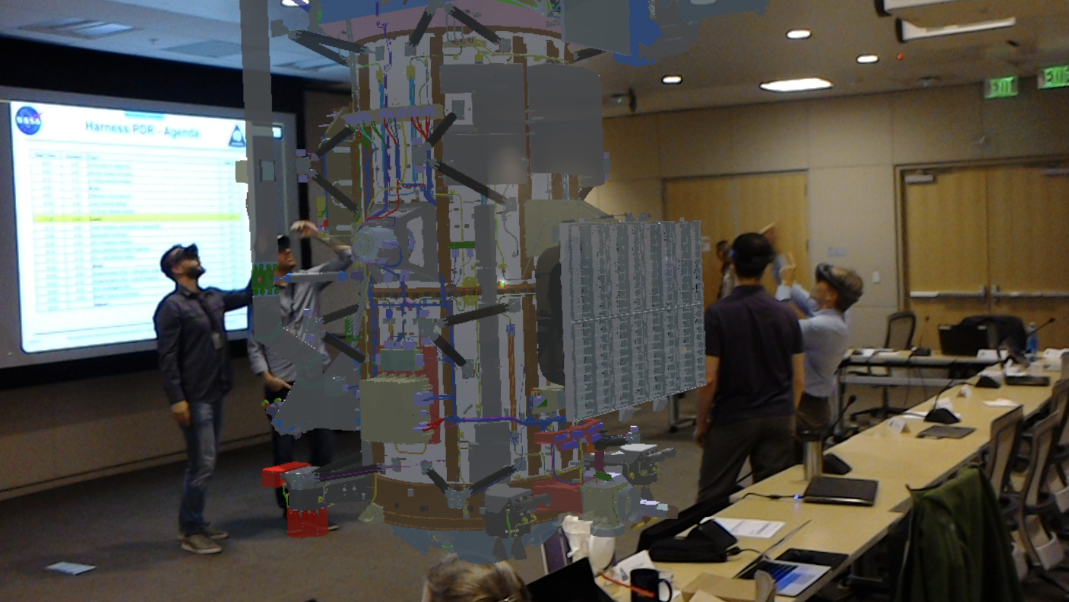}
    \caption{Europa Clipper~\cite{pappalardo2024science} Harness Preliminary Design Review (PDR), where participants utilized ProtoSpace on the HoloLens.}
    \label{fig:clipper-pdr}
\end{figure}

As shown in Figure~\ref{fig:clipper-pdr}, engineers sometimes review and discuss designs of spacecraft using ProtoSpace. We have observed that ProtoSpace can enhance collaborative design reviews by allowing engineers to more confidently and quickly verify (a) overall design and assembly processes; (b) that subsystems and instruments will integrate appropriately; and (c) that existing or potential ground support equipment (GSE) will usable during the integration process. Finally, in some cases, engineers have used ProtoSpace to avoid the need of physical mockups. One engineer noted, “we were going back and forth over email for months, but once we put on the headsets, we immediately knew the answer within a couple of minutes.”

\subsubsection{Integration \& Test, and Assembly Launch and Test Operations}
\begin{figure}
    \centering
    \includegraphics[width=1\linewidth]{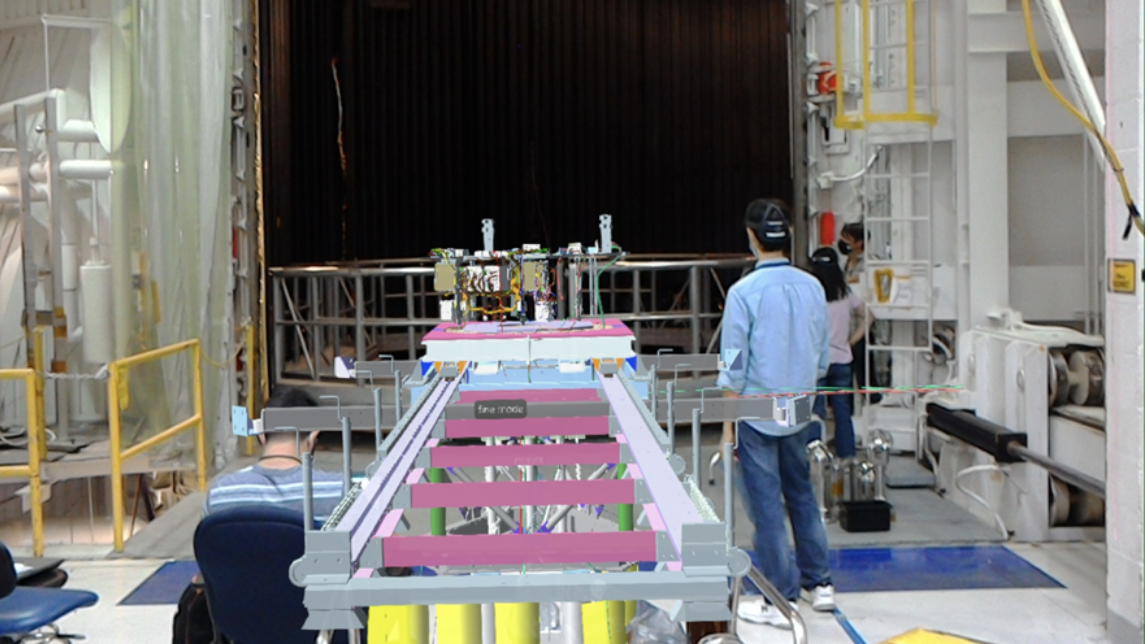}
    \caption{ProtoSpace was used for planning how NISAR~\cite{kellogg2020nasa} would be tested inside the 25ft Thermal Vacuum Chamber (TVAC) at JPL. This included ensuring that all cabling would be accessible.}
    \label{fig:nisar-25ft}
\end{figure}

The second major use case for ProtoSpace occurred during the Integration and Test (I\&T) and Assembly Launch and Test Operations (ATLO) phase. In this use case, ProtoSpace helps teams reach decisions more quickly and with more confidence. In particular, we have observed that engineers could more confidently and quickly perform: clearance checks, verification \& validation of subsystems, reachability analysis, line-of-sight analysis, rehearsing/training for procedures, and verifying integration with facilities \& GSE. In addition, engineers would often plan out where to put equipment, virtually test out GSE before purchasing it, and by doing so, they would often avoid risky or potentially hazardous situations. Figures~\ref{fig:nisar-25ft} shows one example of ProtoSpace being used for test.

\subsubsection{Presentations}
\begin{figure}
    \centering
    \includegraphics[width=1\linewidth]{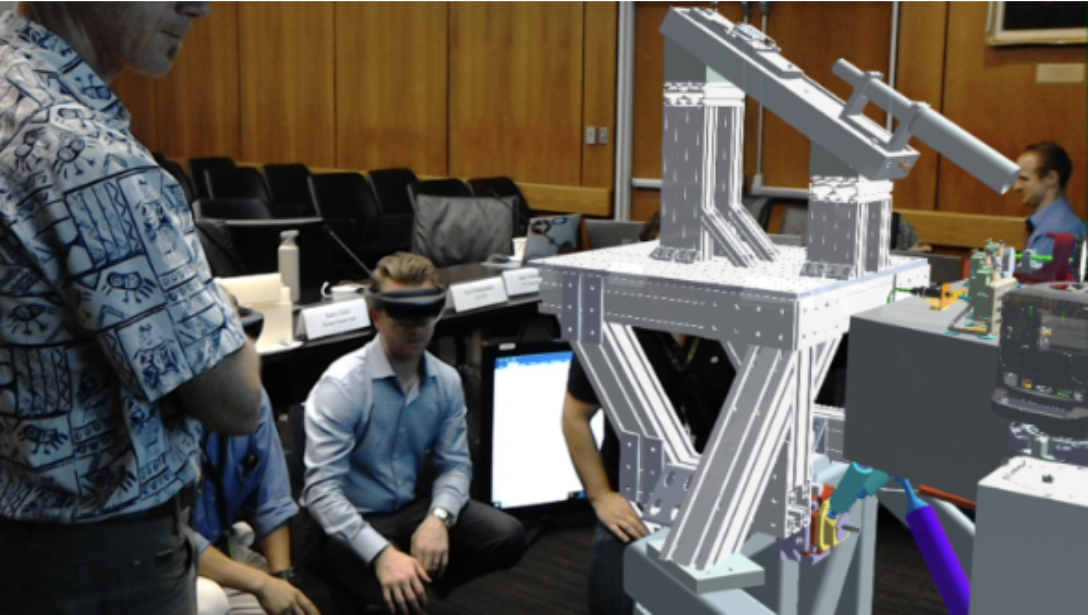}
    \caption{Roman Space Telescope Preliminary Design Review (PDR)}
    \label{fig:roman-pdr}
\end{figure}

\begin{figure}
    \centering
    \includegraphics[width=1\linewidth]{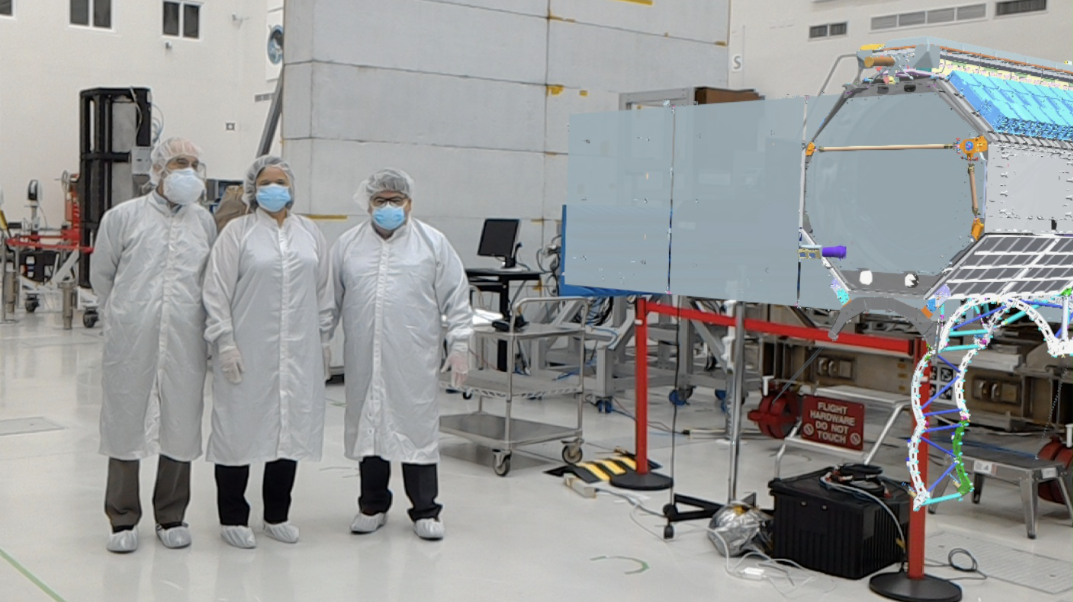}
    \caption{VIP Visitors in the Spacecraft Assembly Facility with a virtual NISAR spacecraft}
    \label{fig:vips}
\end{figure}

\begin{figure}
    \centering
    \includegraphics[width=1\linewidth]{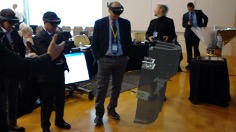}
    \caption{ProtoSpace was used to convey different aspects of the SphereX spacecraft~\cite{crill2020spherex} during a Site Visit Proposal session.}
    \label{fig:spherex-site-visit}
\end{figure}

We have observed that ProtoSpace can help “get the point across” to constituents, stakeholders, and the public. 
Communication is immediately enhanced by bringing the conversation into the 3D spatial realm. 
ProtoSpace has been used for site visit proposals (e.g., Figure~\ref{fig:spherex-site-visit}), hardware design reviews (e.g., Figures~\ref{fig:clipper-pdr} and~\ref{fig:roman-pdr}), VIP visits (e.g., Figure~\ref{fig:vips}), public outreach demos, and meetings with international partners.

\section{System Details}
\label{sec:details}

\begin{figure}
    \centering
    \includegraphics[width=1\linewidth]{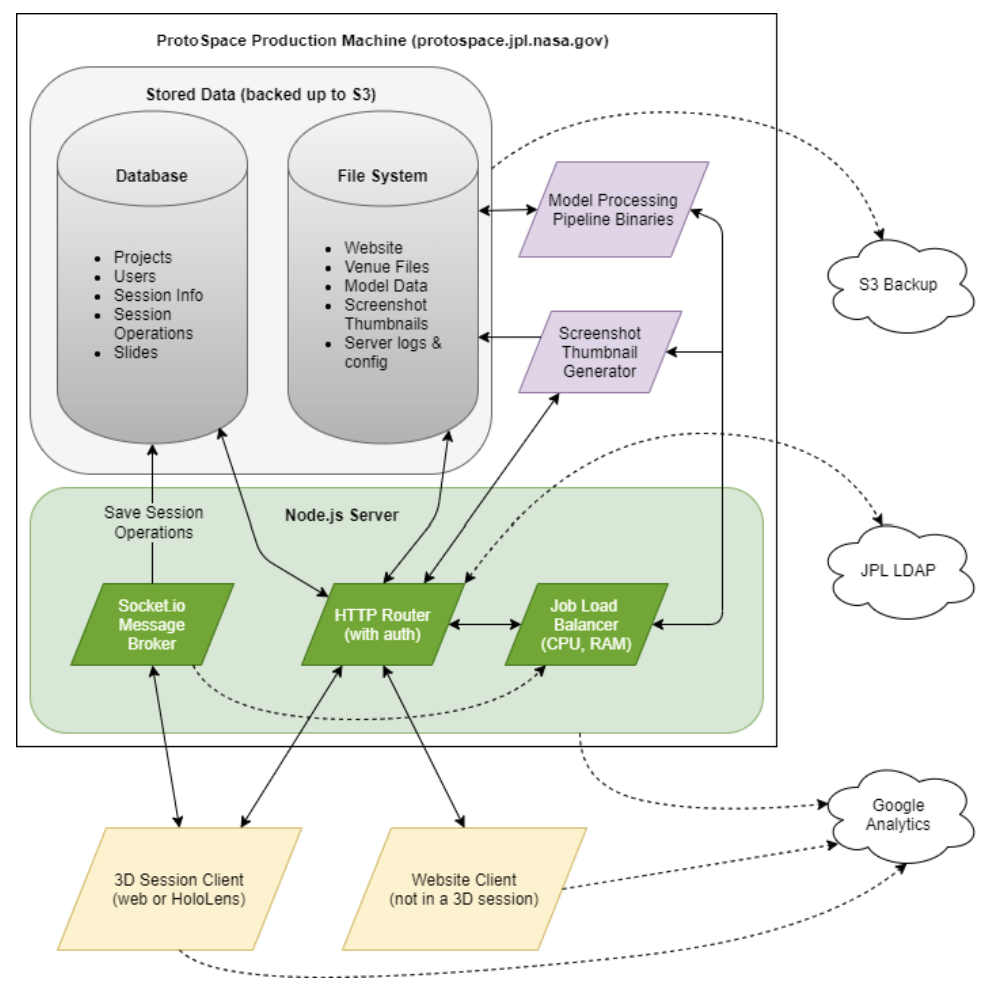}
    \caption{The overall system architecture, from the point of view of the web server backend that can be connected to by either 3D session clients (web or HoloLens) or website clients (not viewing any 3D session).}
    \label{fig:architecture}
\end{figure}

Figure~\ref{fig:architecture} shows the overall architecture for ProtoSpace and the following sections go into detail related to the web server backend (Section~\ref{sec:web-server-backend}), the Hololens client (Section~\ref{sec:hololens-client}), and web portal frontend (Section~\ref{sec:web-portal-frontend}).

\subsection{Web Server Backend}
\label{sec:web-server-backend}
As shown in Figure~\ref{fig:architecture}, the backend server has a variety of functions, including storing housekeeping data (projects, users, session information, etc.) and CAD model data, as well as acting as a message broker between clients to keep session state synchronized.

The ProtoSpace server has a C++ pipeline that imports CAD models and processes them into a proprietary format that is more readily usable in both interactive model reduction as well as client side rendering.
The same pipeline also handles the model reduction process.
As described later on in Section~\ref{sec:web-portal-frontend} and shown in Figure~\ref{fig:reduction}, users interactively choose how to reduce the model's complexity; a quick estimated preview is shown in the browser to the user, but the full processing is handled asynchronously via the C++ pipeline.
The pipeline handles geometry tesselation, generation of levels of detail, etc.
The pipeline relies on a variety of CAD SDK libraries, including the Open CASCADE Technology (OCCT) library, JT Toolkit, Assimp, and the FBK SDK. 
Supported file types include importing .jt, .step, .iges, .fbx, .dae, .3ds, .obj, .ply, and .stl files and exporting to .fbx, .dae, .obj, .stl, and .ply.

The following strategy was employed for persisting ProtoSpace sessions and ensuring that late-joining clients can be immediately synchronized with the current state of the session.
All operations that affect the session---including model manipulation operations, point-of-interest placements, etc.---are sent as network messages to the server.
The server then broadcasts those messages out to all clients, but it also stores the session operations in memory and flushes it to disk for long-term storage.
Such session operations are then ``squashed'' to a smaller set of messages which then represent the persisted, up-to-date session state.
In storing the session state in this fashion, we can then re-use the same message handler functions in each client code base.
This helps avoid potential discrepancies in two separate flows of code for arriving at the same state and also avoids having the server do the work of maintaining a central session state object, which could be complicated when managing piece-part transformations within a CAD model.
However, the network message squashing algorithm must be handled with care as well to avoid bugs.
ProtoSpace slides are also implemented by storing the current set of session operations.

Another function the ProtoSpace server performs asynchronously is screenshot thumbnail generation for all models, sessions, and slides.
Screenshot generation is performed by launching a web client renderer (see Section~\ref{sec:web-renderer}), potentially in headless mode, that loads the model, session, or slide, and then rotates the virtual camera view around the model taking screenshots at each perspective.
A configurable discrete number of viewpoints are used for this purpose.
Finally, the screenshots are stitched together into a sprite sheet for use as rotating animation driven by mouse move in the web portal frontend, providing users with a preview of the content.

The ProtoSpace server has been tested running on a single Windows workstation with up to 20 HoloLens devices simultaneously connected in a single session.
A microservices architecture was considered and developed early on, but was eventually discarded as unnecessary.

\subsection{HoloLens Client}
\label{sec:hololens-client}
This is the main ProtoSpace application (``app'' or ``client'') for HoloLens 1 and 2.  
Each participant in a ProtoSpace session wears a HoloLens that is running the ProtoSpace HoloLens client.  
The client app handles all of the major ProtoSpace functions including model download and storage, real-time model rendering, the interactive toolbox, and real-time synchronization of session state (e.g., the location of the point-of-interest marker, if any) with other clients in the session.

Typically a ProtoSpace session will have from one to ten or more HoloLens clients participating.  
The limitation on the total number of clients in the session is typically the number of available HoloLens devices and the physical logistics of fitting people into a space.  
Note that the more people crowd together in a given space the more likely it will be that they will occlude each others' HoloLens' view of the space; each HoloLens must be able to see parts of the fixed space (walls, ceiling, furniture, etc) to maintain its understanding of where it is.

All HoloLens clients in the session communicate with each other in real time via the ProtoSpace Server (see Section~\ref{sec:web-server-backend}).  
This is how they maintain synchronization of session state.  
HoloLenses may leave and join a particular live session at any time.

HoloLens clients also access the ProtoSpace Server as needed to download models.  
Note that once a model has been downloaded on a particular HoloLens it is persistently cached in the local flash memory of that device and will typically not be re-downloaded unless it is updated on the server.

We now describe in more detail five important aspects of the HoloLens client: the user interface design, the toolbox, alignment, and rendering.

\subsubsection{User Interface Design}

The main design of the ProtoSpace HoloLens user interface took into consideration the fact that many of original targeted end users were usually familiar with existing CAD tools.
Thus, they may be familiar with 3D manipulation tools such as rotation, translation, scale, hide/show, and section planes.
Most CAD GUIs are comprised of rectangular panels containing modal button tool sets, framing a Euclidean, non-stereo, 3D workspace.  In CAD, this design makes the tools accessible while minimizing occlusion of the 3D model.  However, in ProtoSpace the models are visualized in physical space, as mixed reality ``holograms,'' and the user is constantly moving, so if the tool interface is locked to a particular location the user will have difficulty accessing it at all times, and it may occlude view of the model at other times.
An additional constraint for ProtoSpace is the HoloLens display's field of view. 
Version 1 was limited to approximately 30 degrees x 17.5 degrees, while the HoloLens 2 was approximately 44 x 29.5 degrees. 
This limitation makes it extremely important to use display real estate efficiently.  
The ProtoSpace ``Toolbox'' user interface is designed to be there when it is needed, and get out of the way when it is not.

To achieve these qualities, we designed novel features into the interaction system~\cite{clausen2020collaborative}.  
These include:
\begin{itemize}
    \item Menu minimization and sub-menu transformation
    \item ``Follow and Catch'' behavior (based on users' head gaze acting as a ``cursor'')
    \item Continuity of motion
    \item Obstruction avoidance
    \item Maintain perceived distance between User and model
    \item Constrained versus Unconstrained manipulation
    \item An iconographic menu
\end{itemize}

\subsubsection{Toolbox}

\begin{figure*}[h]
    \centering
    \begin{subfigure}[t]{.2\linewidth}
      \centering
      \includegraphics[width=.95\linewidth]{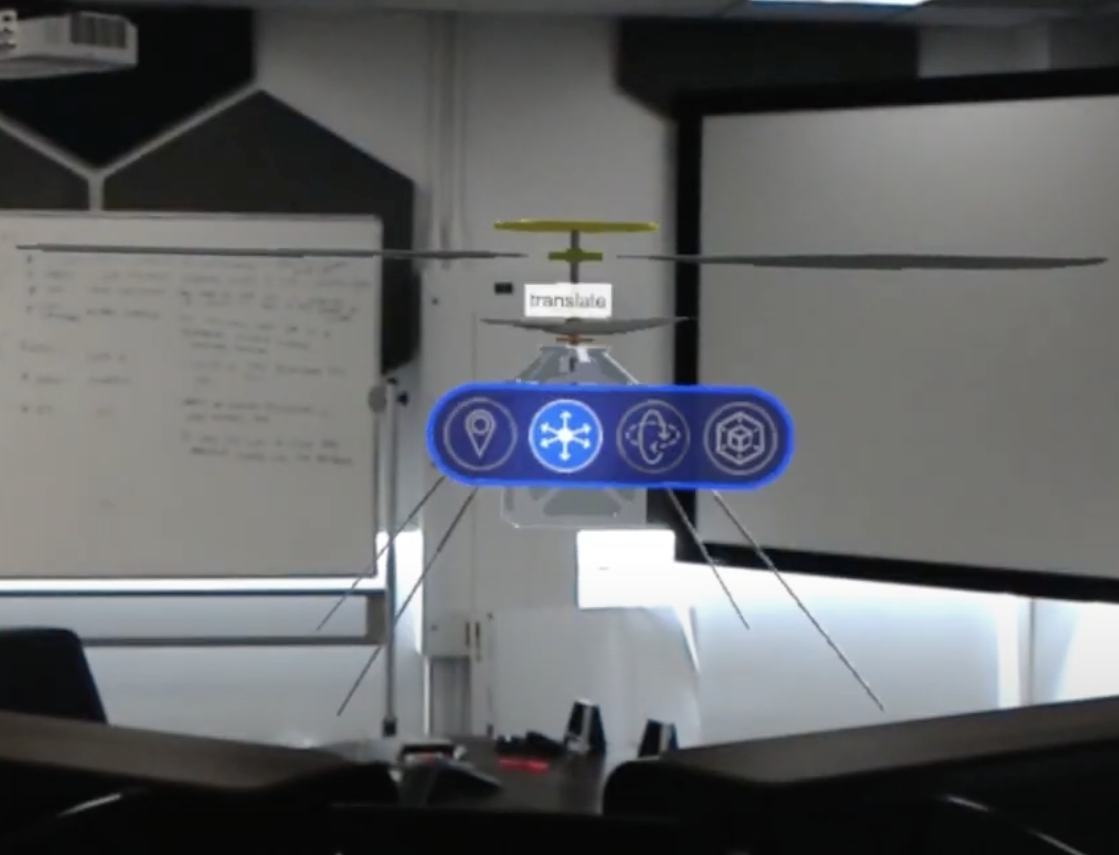}
      \caption{\centering The Move tool\\is highlighted.}
    \end{subfigure}%
    \begin{subfigure}[t]{.2\linewidth}
      \centering
      \includegraphics[width=.95\linewidth]{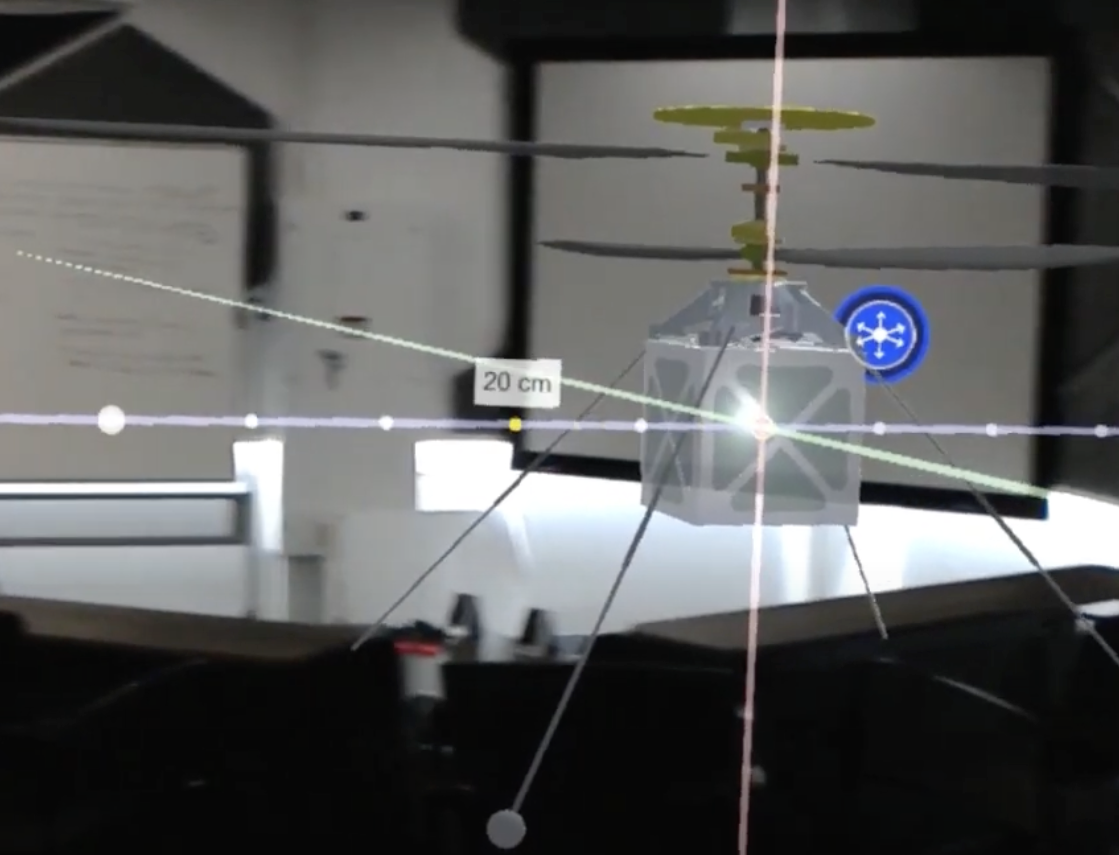}
      \caption{\centering X,Y, and Z axes aid\\in precision movement.}
    \end{subfigure}%
    \begin{subfigure}[t]{.2\linewidth}
      \centering
      \includegraphics[width=.95\linewidth]{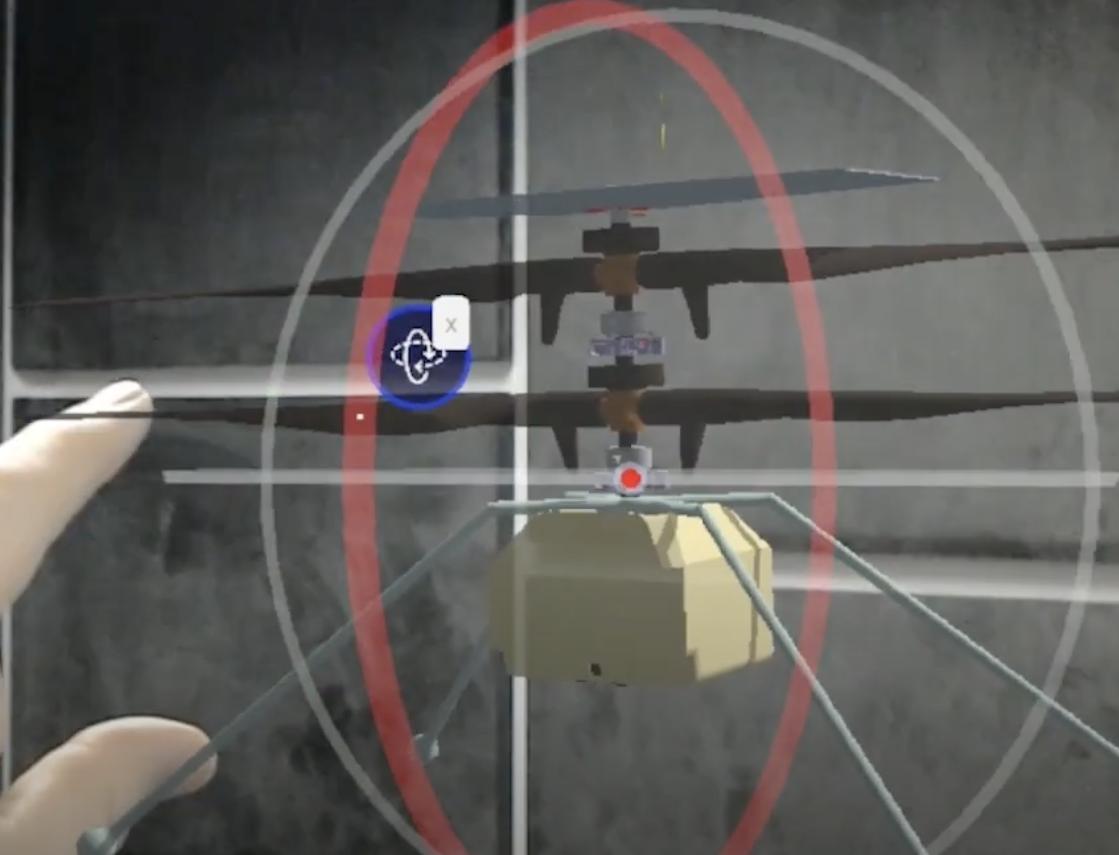}
      \caption{\centering The Rotate tool's\\ circular axes widgets.}
    \end{subfigure}%
    \begin{subfigure}[t]{.2\linewidth}
      \centering
      \includegraphics[width=.95\linewidth]{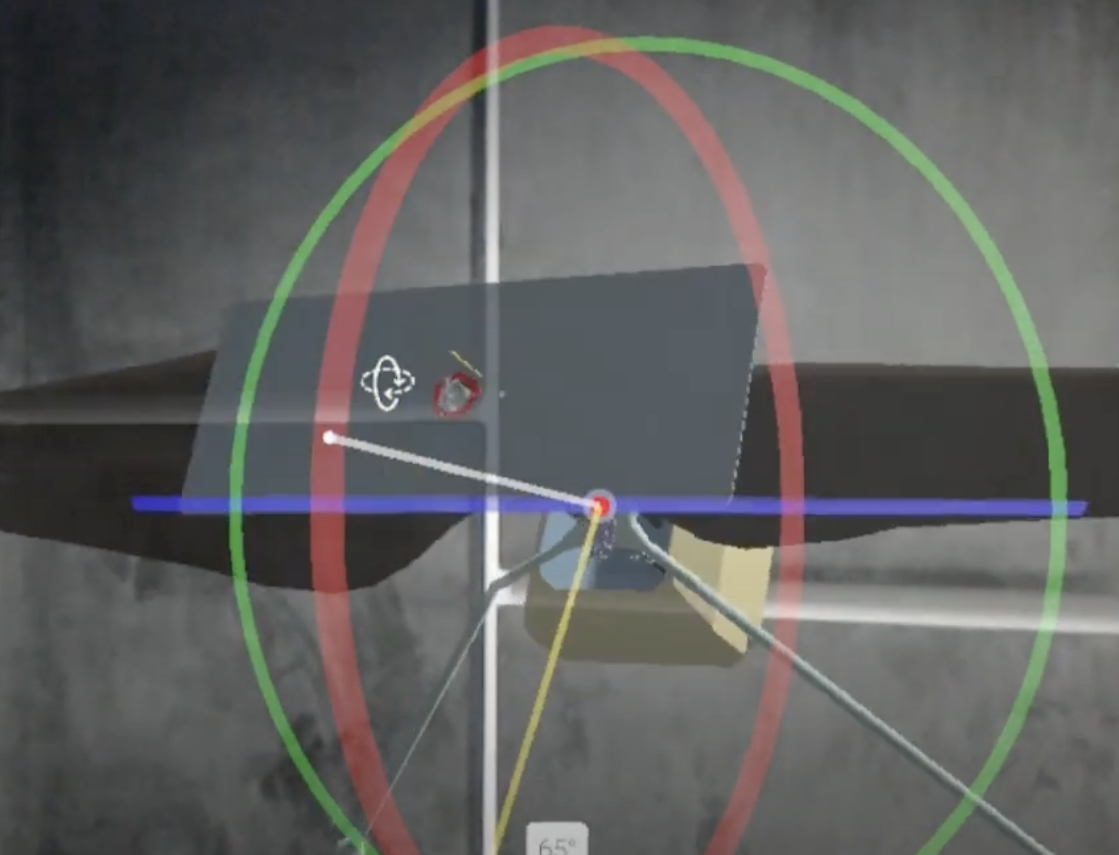}
      \caption{\centering Rotating on\\the x-axis.}
    \end{subfigure}%
    \begin{subfigure}[t]{.2\linewidth}
      \centering
      \includegraphics[width=.95\linewidth]{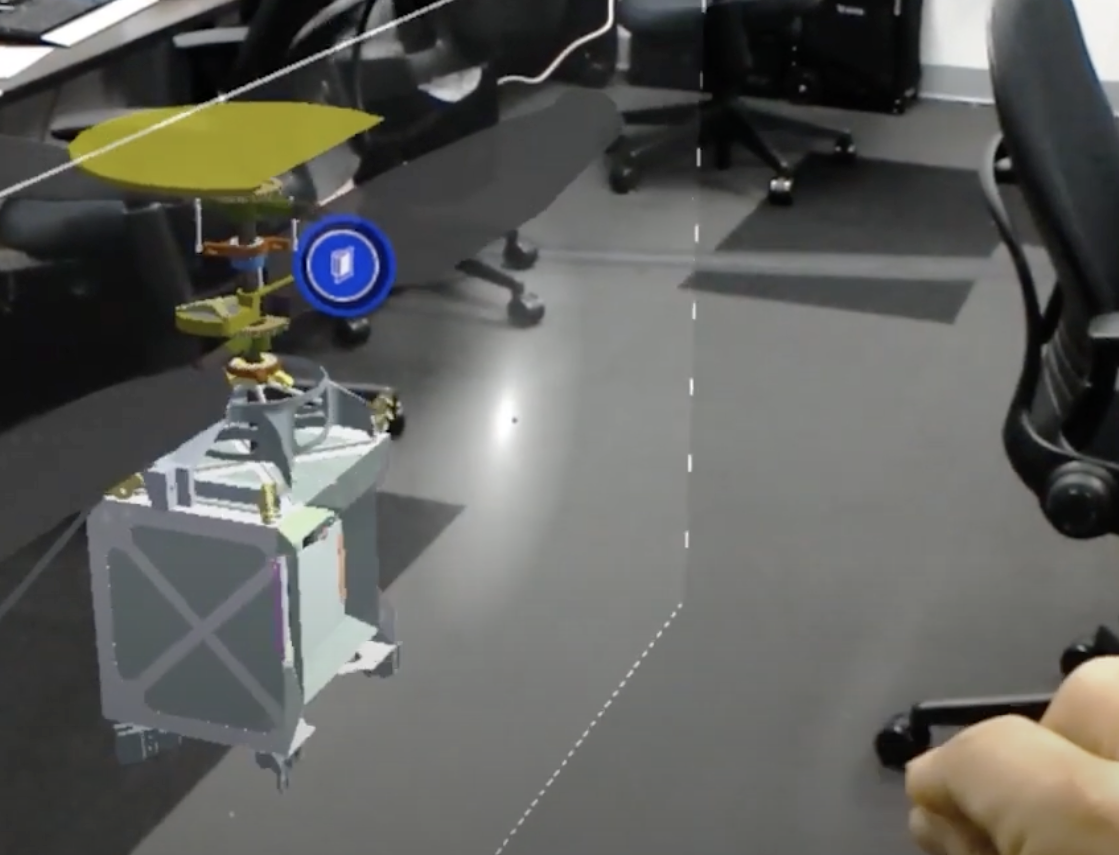}
      \caption{\centering The Cut plane tool.}
    \end{subfigure}%
    \caption{The move, rotate, and cut plane tools in ProtoSpace shown with the Mars Ingenuity Helicopter~\cite{balaram2021ingenuity}. The corresponding YouTube video~\cite{protospaceYouTube} shows these tools in video format.}
    \label{fig:move-rotate-cutplane}
\end{figure*}

\begin{figure*}[h]
    \centering
    \begin{subfigure}[t]{.25\linewidth}
      \centering
      \includegraphics[width=.95\linewidth]{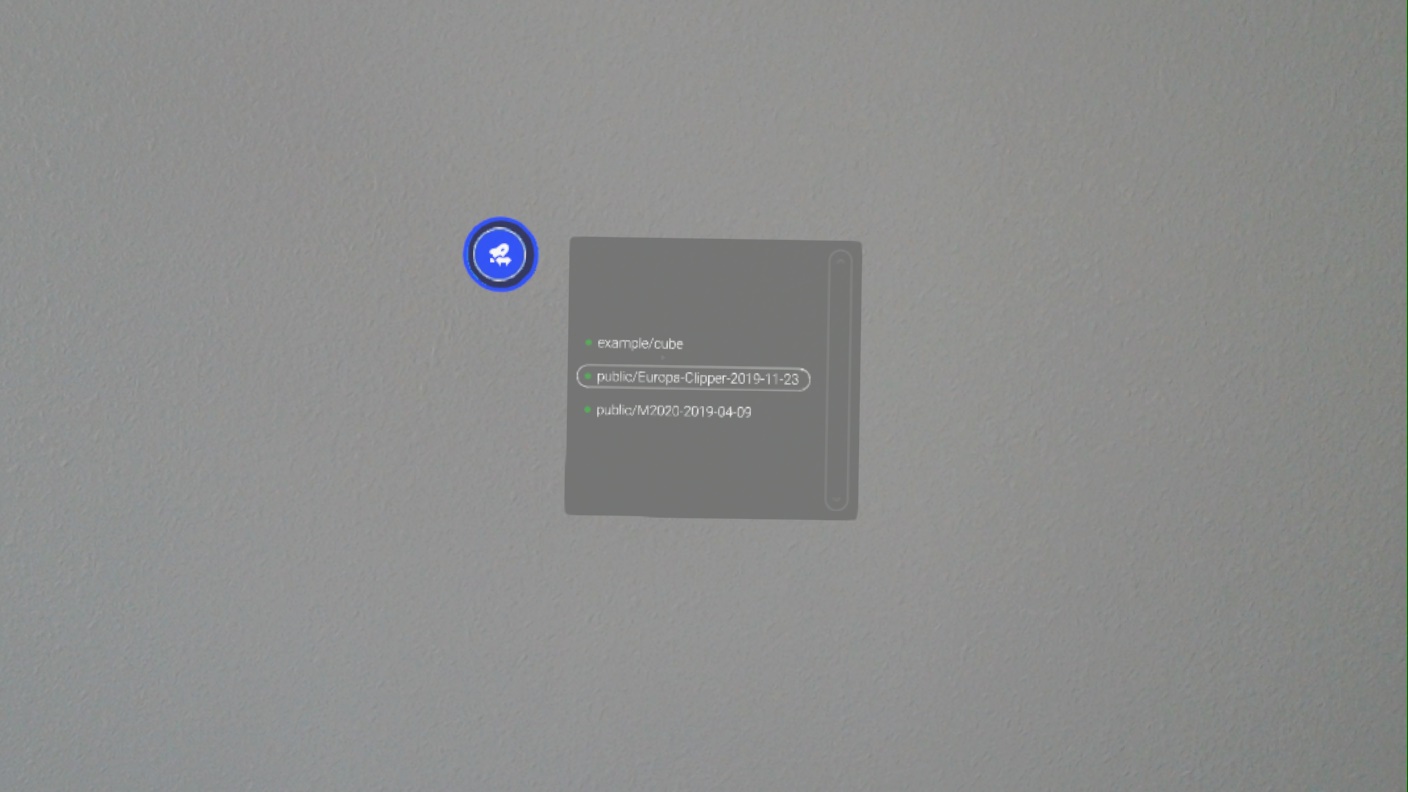}
      \caption{\centering The original model selection interface.}
    \end{subfigure}%
    \begin{subfigure}[t]{.25\linewidth}
      \centering
      \includegraphics[width=.95\linewidth]{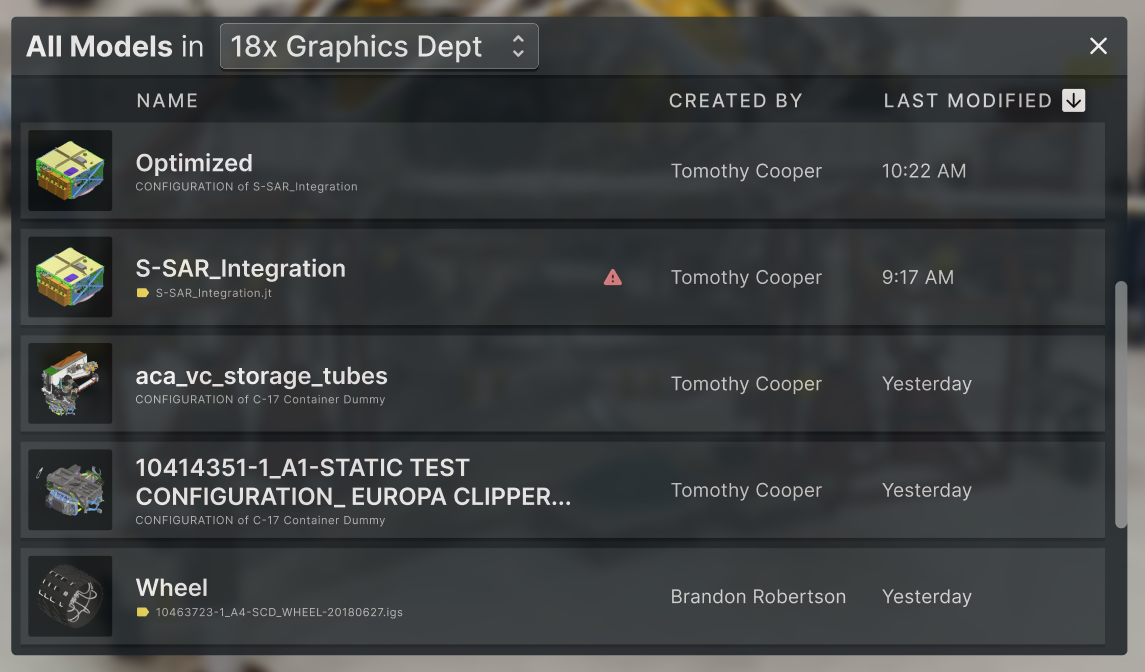}
      \caption{\centering The model selection interface.}
    \end{subfigure}%
    \begin{subfigure}[t]{.25\linewidth}
      \centering
      \includegraphics[width=.95\linewidth]{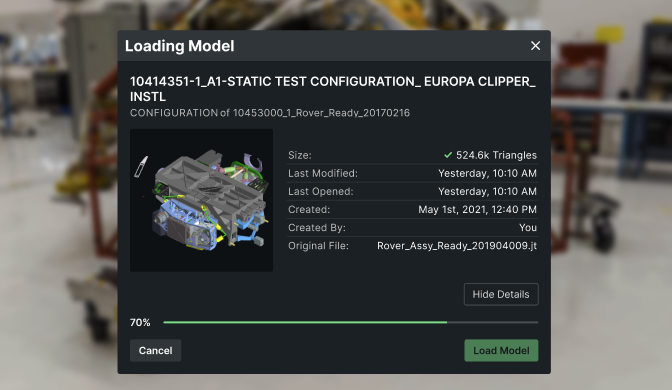}
      \caption{\centering The model confirmation screen.}
    \end{subfigure}%
    \begin{subfigure}[t]{.25\linewidth}
      \centering
      \includegraphics[width=.95\linewidth]{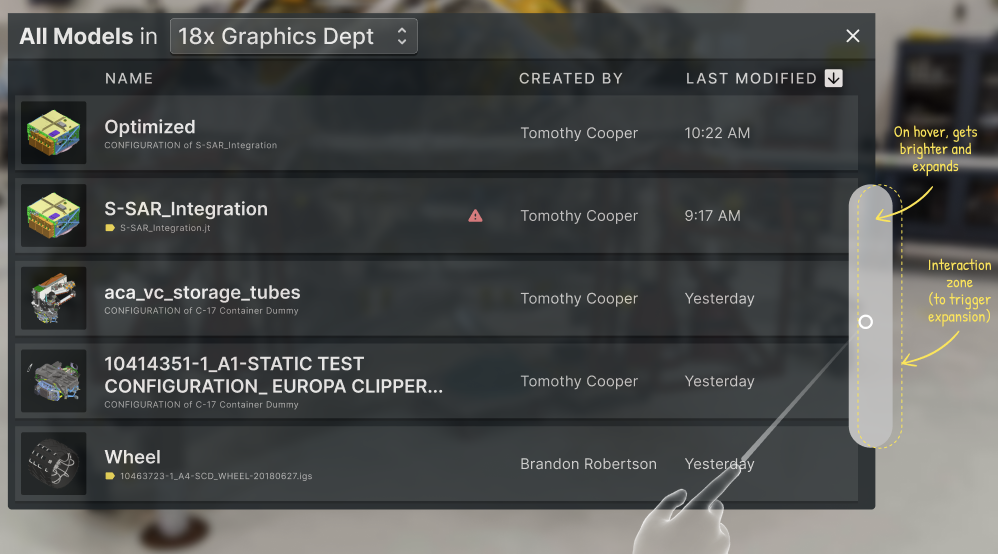}
      \caption{\centering The custom scroll bar.}
    \end{subfigure}%
    \caption{The model selection interfaces of ProtoSpace.}
    \label{fig:model-selection-tool}
\end{figure*}

The following are a subset of the tools created for ProtoSpace over the years:
\begin{itemize}
    \item POI --- place a point-of-interest with an air-tap; this appears as a 3D chevron that is shared between HoloLens clients in the same session. Upon placement, a sound effect is triggered to gain the users' attention. If outside the field of view on a headset, another chevron appears at the edge of the display to indicate to the user where to look.
    \item Model Manipulation Tools (move, rotate, and scale allowed for whole model manipulation or piece-part manipulation):
    \begin{itemize}
        \item Move --- see Figure~\ref{fig:move-rotate-cutplane}; the user can either freely move the model (or part) unconstrained, or via constrained movement along one of the three predefined X, Y, or Z axes.
        \item Rotate --- see Figure~\ref{fig:move-rotate-cutplane}; the user can either free rotate the model (or part) unconstrained, or via constrained movement along one of the three predefined X, Y, or Z axes.
        \item Scale --- the user can scale to preset or arbitrary sizes
        \item Cut Plane tool --- see Figure~\ref{fig:move-rotate-cutplane}; an air-tap dragging gesture moves the cut plane along one of the predefined axes (X, Y, or Z); the sub-menu for this tool allows the user to switch which axis is being used
        \item Animation tool --- play controls for animations
    \end{itemize}
    \item Primitive --- for authoring simple 3D objects
    \item Procedures --- this opens an entirely different user interface for working through a procedure; see Section~\ref{sec:procedures} for more details.
    \item Model select menu --- See Figure ~\ref{fig:model-selection-tool}; Unity primitives were originally used for the model selection interface. This interface was later updated with a custom design using interactive components from Microsoft Mixed Reality Toolkit 2 \cite{mrtk2} along with custom implementations for an interactive and expandable scroll bar and text-ticker for long-form text viewing. Users may filter between projects when selecting models.
    \item Alignment tool --- for adjusting the HoloLens' origin coordinate frame
    \item Slides --- for creating, deleting, and switching between different slides
\end{itemize}

\subsubsection{Alignment}
There are two reasons we need to do alignment:
\begin{enumerate}
    \item Virtual-to-physical alignment --- where we are concerned with precisely aligning a virtual model to a physical environment. This could be where one needs to see how physical hardware (virtually shown in ProtoSpace) will fit with respect to its surroundings (e.g., for movement of spacecraft hardware through a narrow corridor). Or, this could be for when one wants to see how the next part of hardware (virtually shown in ProtoSpace) will be assembled with already existing hardware.
    \item Multiple headset alignment --- where multiple users wearing XR headsets need to all share the same coordinate frame so they see the virtual content in the same location. Typically, when a HoloLens is started, its initial origin coordinate frame is positioned at that starting location; thus, there is a need to align all the individual coordinate frames.
\end{enumerate}

While a variety of methods were initially investigated to achieve alignment, here we report on two approaches that worked well.
While one user may be able to, through gesture-based manipulation via the move and rotate tools, precisely achieve alignment, it is hard to repeat this process between sessions as well as between headsets.
Thus, marker-based alignment offers a repeatable approach to achieve alignment.
If the accuracy of the alignment is not good enough, fine-tuning may occur through minor adjustments on a per-headset basis.

\begin{figure}
    \centering
    \begin{subfigure}[t]{.33\linewidth}
      \centering
      \includegraphics[width=.95\linewidth]{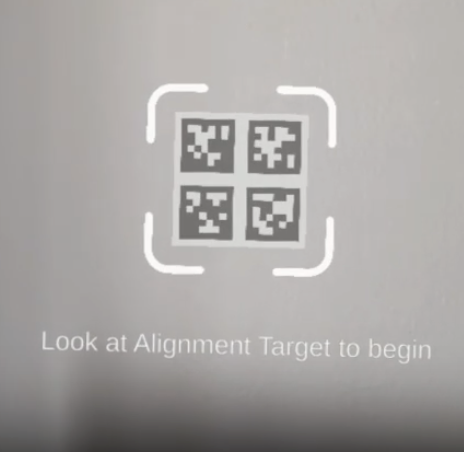}
      \caption{Initial view}
    \end{subfigure}%
    \begin{subfigure}[t]{.33\linewidth}
      \centering
      \includegraphics[width=.95\linewidth]{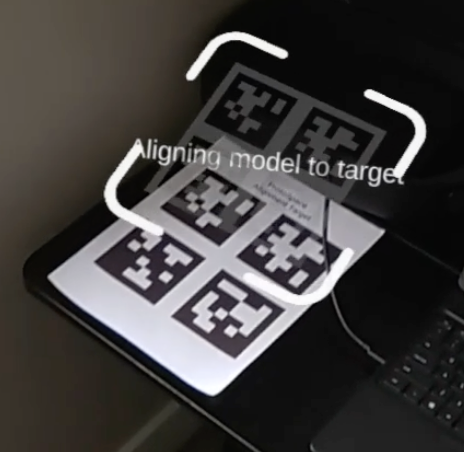}
      \caption{\centering Animates to \\the marker}
    \end{subfigure}%
    \begin{subfigure}[t]{.33\linewidth}
      \centering
      \includegraphics[width=.95\linewidth]{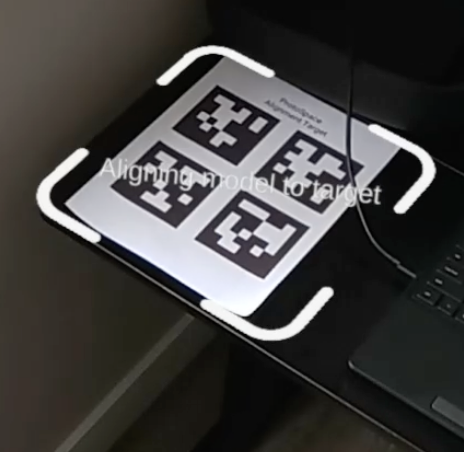}
      \caption{\centering Aligned to \\the marker}
    \end{subfigure}%
    \caption{ProtoSpace alignment target using 4 AprilTags~\cite{olson2011apriltag}. Here, we see the general sequence of the user experience -- In (a), the user is shown a virtual marker of 4 AprilTags with instructions, ``Look at Alignment Target to begin.'' In (b), as the user points the HoloLens' view towards the physical alignment target, the virtual marker animates towards the physical tag while also fading out. Finally, in (c), the four corner outline of the visualization pulsates in and out to indicate to the user that alignment is taking place.}
    \label{fig:alignment}
\end{figure}

First, we used marker-based alignment via a set of AprilTags. 
Early approaches used two or three separate small AprilTags (e.g., around 1-2 inches in size), positioned far apart to minimize alignment errors that could more easily occur with using a single marker; one example is shown in Figure~\ref{fig:cal-procedure}.
However, we ultimately settled on one larger set of 4 AprilTags as shown in Figure~\ref{fig:alignment} that can be easily printed on standard 8.5x11 inch paper.
Four AprilTags were used so that we could detect up to 20 points (4 corners and 1 center point, for each four tags), thus mitigating any potential occlusion and increasing potential alignment accuracy.
The final alignment calculation was done via a Perspective-n-Point algorithm using OpenCV.
Finally, if the user can place the AprilTag alignment marker directly onto the physical asset for which there is a virtual CAD model loaded into ProtoSpace, the web client 3D session viewer allows for a user to place a virtual AprilTag for quicker session set up; see Figure~\ref{fig:april-tags-side-by-side}.

\begin{figure}
    \centering
    \includegraphics[width=1\linewidth]{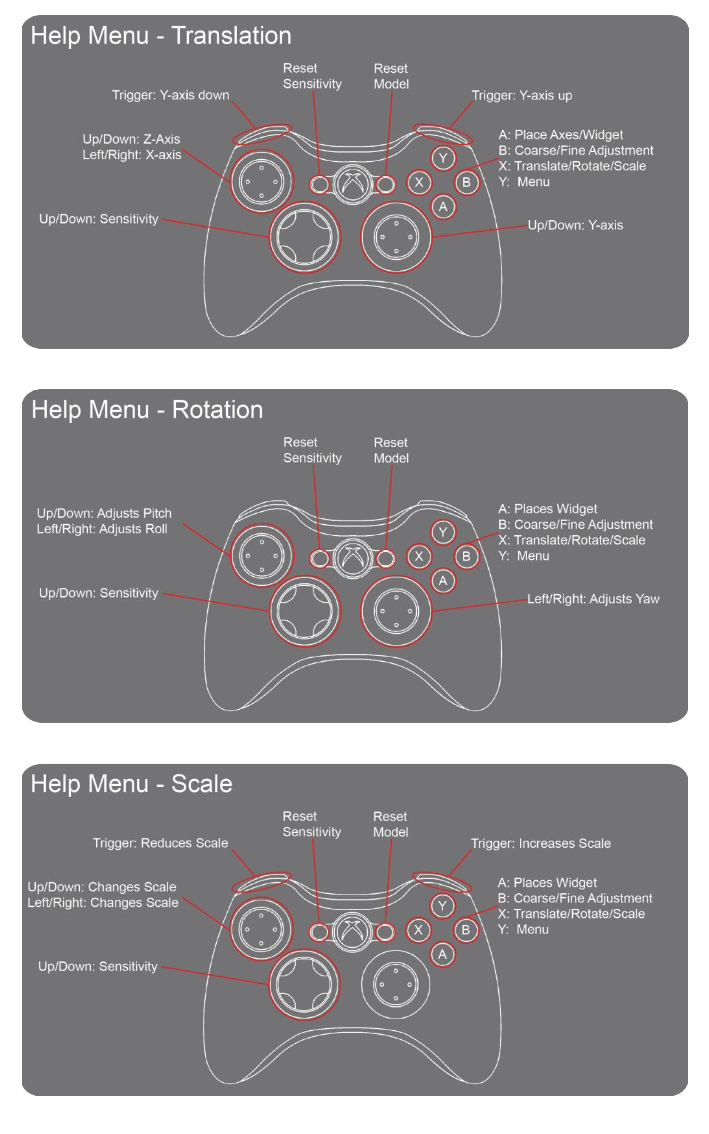}
    \caption{XBox One controller mapping to ProtoSpace}
    \label{fig:xbox-controller-mapping}
\end{figure}

Second, we also implemented an XBox Controller approach for alignment.
In use cases, such as SWOT, where a user needs to align a CAD model to a real world object, rotating, translating, and scaling using the HoloLens' tap and drag gesture provides a less-than-ideal method for alignment. 
An XBox One controller was mapped to ProtoSpace's translate, rotate, and scale functionality to overcome this challenge and provide fine control of the CAD model's orientation in augmented reality. The controller mapping for each feature is shown in Figure \ref{fig:xbox-controller-mapping}. The direction of translation was intuitively mapped to be in relation to the camera's (i.e. the user's) perspective; similarly, rotation and scale are in reference to the User's gaze on the CAD model.

\subsubsection{Rendering}
\label{sec:rendering}
As previously mentioned in Section~\ref{sec:related-work-xr-cad-viewing}, a key challenge for XR CAD viewing is how to render massive CAD models on untethered headsets.
For ProtoSpace, the key insight was to allow users to interactively reduce the model's complexity beforehand in a simple web interface (see the following sections).
Unfortunately, sometimes it is not fully possible to reduce the model complexity enough to get it rendering at high frame rate through naive rendering; for example, it could be the user wants to keep the model's tesselation accuracy very high.

ProtoSpace employs a number of strategies to further optimize rendering performance (many of which have been employed by other prior work~\cite{baxter2002gigawalk}).
This includes doing occlusion and frustum culling in clever ways (e.g., in a multi-threaded approach, and processing occluding nodes based on their rasterization size); level-of-detail (LOD) rendering; stereo instanced mesh rendering; using dithered transparency (to avoid multiple rendering passes); packing mesh attributes; etc.
Additionally, ProtoSpace also cached models on disk and in memory, which avoided the need to download models from the ProtoSpace server and also made switching between models fast during a single session.

We also make note here of our experience with using the HoloLens reprojection feature~\cite{hologramStability}.
ProtoSpace projects a generic dot pattern to find an average depth of the visible scene, and based on this sets the HoloLens stabilization plane at that depth, with the plane normal parallel to the headset gaze direction.
At one point, we tried the ``automatic planar reprojection'' provided by a newer HoloLens version, but it gave visually unpleasing results---specifically the plane's normal appeared to be not parallel to the gaze direction and this looked like a strong ``shearing'' artifact when rendering the Mars Perseverance rover.
It could be that this was due to the fact that from certain perspectives, parts of the model might look close and other parts far away, all within a similar viewing angle.
Another potential (or exacerbating) cause could be that the Mars Perseverance rover would render under 30 frame per second in some instances during our testing.

Finally, we note that ProtoSpace also has the ability to render ``occlusion-only nodes'' in the CAD model.
This is where certain parts of the CAD model are rendered to the depth buffer only so that they're not visible but still occluded other parts of the rendered model.
This helps for situations where the physical model is only partially built and one wants to render the virtual model in augmented reality; by setting the virtual nodes that correspond to the physical model as ``occlusion-only'' nodes, the physical model appears to occlude the rendered virtual model.
This was especially useful for situations where harness engineers wanted to see the cable harness around a physical model.
This feature was originally built for the NISAR mission (see also Section~\ref{sec:nisar}).

\subsection{Web Portal Frontend}
\label{sec:web-portal-frontend}

The web portal frontend was designed to support HoloLens sessions by allowing users to perform functions on the web that would otherwise be difficult or impossible in the HoloLens. Through the portal, users can manage and interactively optimize models, create and participate in sessions, and perform administrative actions. The portal is built on React, Polymer, and Three.js.

\subsubsection{UI Overview} 
The following are the main pages available in the UI:
\begin{itemize}
    \item Models page --- the user can upload and view the processing status of CAD models as well as export uploaded models to other formats.
    \item Sessions page --- the user can create, join, and delete collaborative sessions.
    \item Admin page --- the user can manage permissions of system users as well as manage \textit{projects} which are used to organize models and sessions.
    \item Model reduction page --- the user can load a CAD model in 3D and perform operations to optimize the viewing performance of the model. See the following Section~\ref{sec:model-reduction} for more details.

\end{itemize}

\subsubsection{Model Reduction}
\label{sec:model-reduction}
\begin{figure}
    \centering
    \includegraphics[width=1\linewidth]{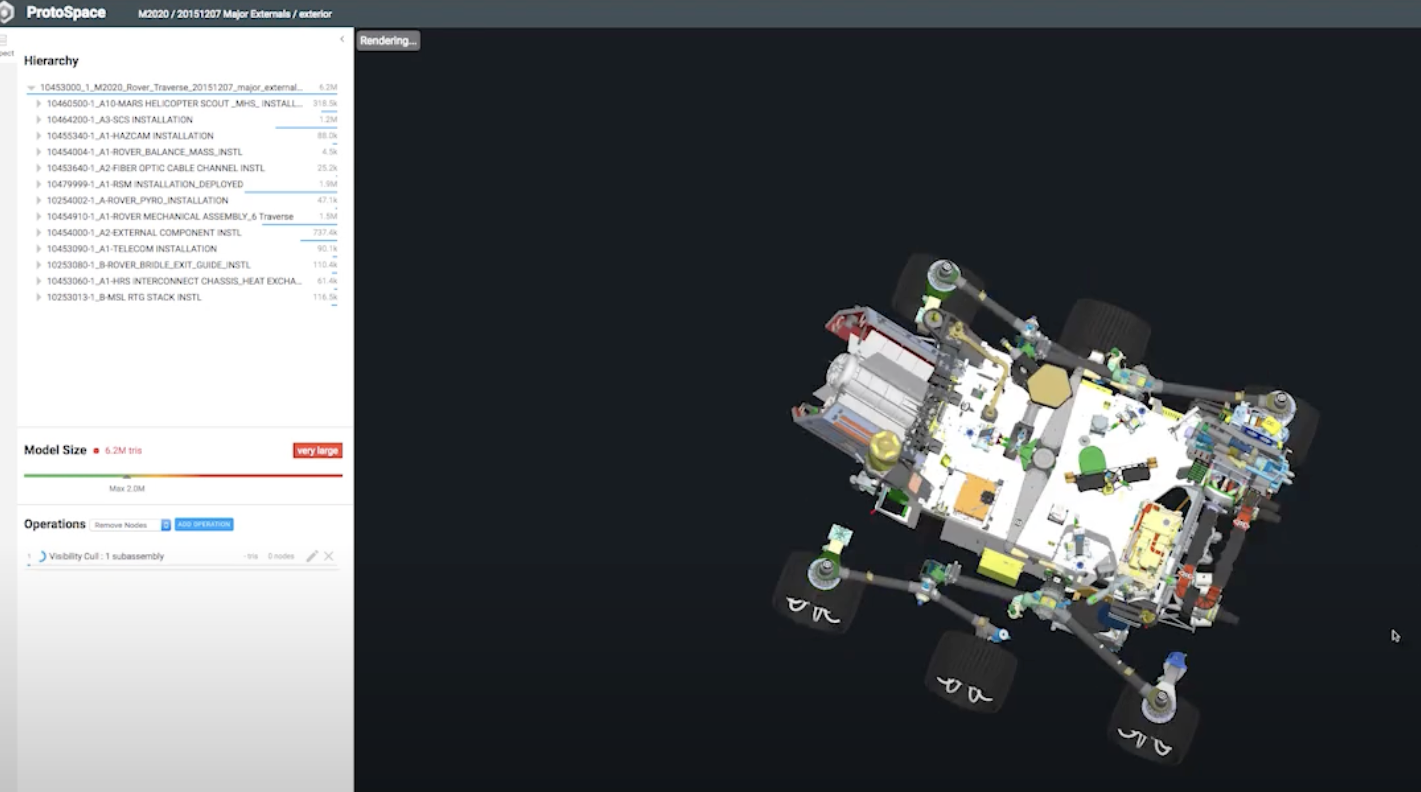}
    \caption{The ProtoSpace model reduction editor provides an interactive way for users to selectively modify the model to reduce its complexity for rendering on XR headsets.}
    \label{fig:reduction}
\end{figure}

Users must perform optimizations on large CAD models through the portal in order for the models to be viewable in the HoloLens. See Figure~\ref{fig:reduction}. The user is provided with a target number of triangles that they should aim to meet to guarantee adequate performance in the HoloLens (configurable thresholds are set by default to 2 million triangles as the ideal count, and 3 million triangles as the secondary limit). Through the model reduction interface, users can create an ordered list of reductions and view a 3D preview of the resulting model. The available reductions include the following:
\begin{itemize}
    \item Remove Nodes --- the user can select any number of nodes to remove from the model.
    \item Remove Nodes by Size --- the user can remove all nodes smaller than a given size in meters.
    \item Remove Nodes by Name --- the user can remove nodes by name or regular expression.
    \item Remove Nodes by Type --- the user can remove nodes by type if provided by the metadata in the CAD file.
    \item Visibility Cull --- the user can remove details not visible to the viewer by creating a sphere outside the model, placing an array of virtual cameras inside the sphere, and then only keeping nodes that are visible to the cameras. This operation is best for optimizing exterior views as interior nodes will be discarded.
    \item Box Cut --- the user can draw a box and cut out or retain the geometry intersecting the box.
    \item Set Color of Nodes --- the user can set the color of nodes.
    \item Set Opacity of Nodes --- the user can set the opacity of nodes.
    \item Set Nodes to Occlusion Only --- the user can set nodes to occlude other elements without actually being displaying, which is useful when overlaying a virtual model on top of a real-world platform. For example, a user might set nodes to occlusion only when displaying a virtual wiring model on top of a real spacecraft chassis in order to prevent wires on the back of the virtual chassis from showing through the real-world chassis.
\end{itemize}
Once the user has completed their reductions, the C++ pipeline on the server will apply the reductions and output an optimized model for use in sessions.

\subsubsection{Session Viewer} 
\begin{figure}
    \centering
    \includegraphics[width=1\linewidth]{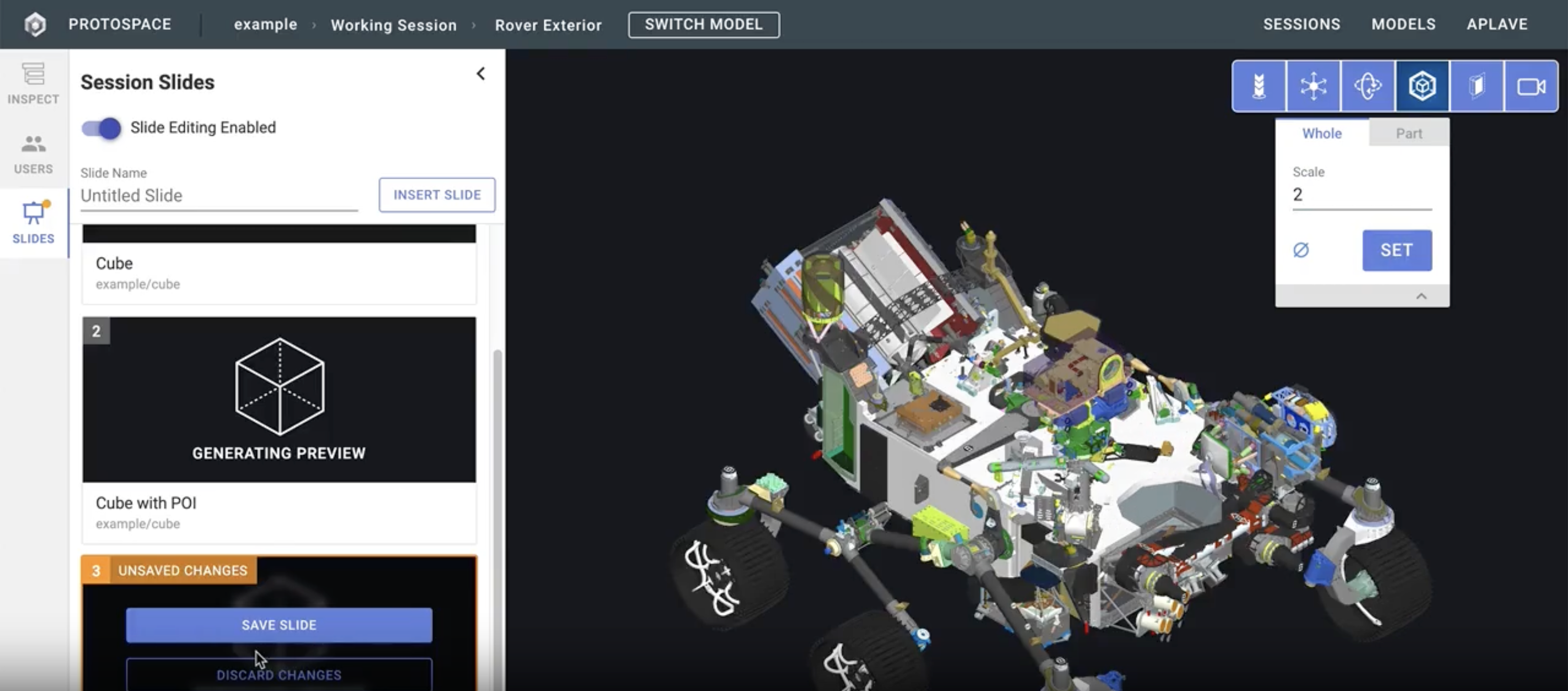}
    \caption{The 3D web session viewer provides an easy was to utilize ProtoSpace's ``slides'' feature that stores a snapshot of the sessions state at a given time.}
    \label{fig:slides}
\end{figure}

\begin{figure}
    \centering
    \includegraphics[width=1\linewidth]{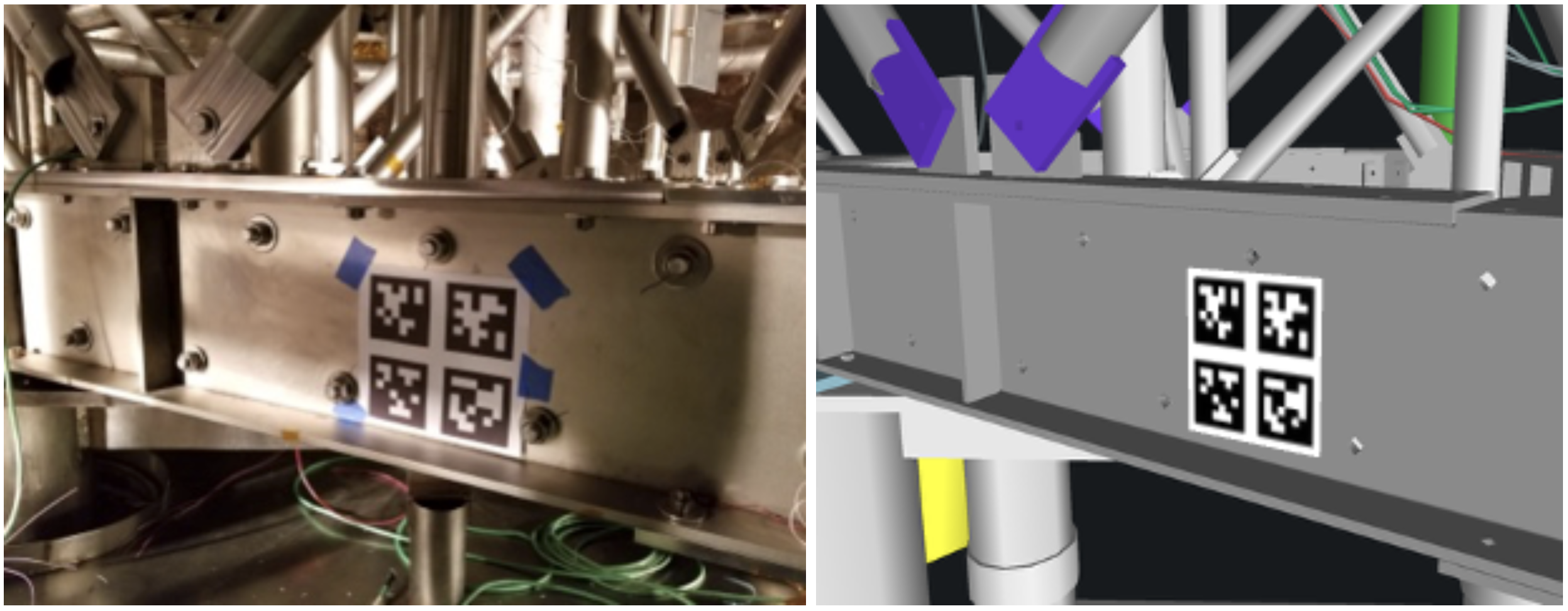}
    \caption{The 3D web session viewer allows for the virtual placement of ProtoSpace’s Alignment Marker (right) so that the virtual model is lined up with where the physical Alignment Marker is placed (left).}
    \label{fig:april-tags-side-by-side}
\end{figure}

The portal allows users to join collaborative real time model viewing sessions on the web. Like HoloLens clients, web clients communicate with other web and HoloLens clients in real time via the ProtoSpace Server in order to maintain synchronization of session state. The session viewer allows users to more easily perform tasks that would be difficult in the HoloLens and additionally allows users without HoloLenses to participate in collaborative sessions. The following is a list of capabilities found in the session viewer:
\begin{itemize}
    \item View Model Hierarchy --- the user can view hierarchy of the nodes in the model as well as highlight and control the visibility of nodes.
    \item Manage Session Users --- the user can view the list of users who have joined the session.
    \item Place Virtual AprilTag --- the user can place a virtual AprilTag within the session to aid in aligning HoloLens user headsets. See Figure~\ref{fig:april-tags-side-by-side}.
    \item 3D Renderer --- the user can view the model in 3D, transform the model with HoloLens-like tools, and place points of interest on the model. See the following Section~\ref{sec:web-renderer} for more details.
    \item Virtual HoloLens Avatars --- 3D HoloLens models are displayed for each HoloLens participant in order for web users to see where the HoloLens users are positioned and where they are looking. 
    \item Slides --- users can create and load snapshots of the session in order to easily present different configurations, views, and models within a session. See Figure~\ref{fig:slides}.
    \item Switch Session Model --- users can switch the active model in a session for all clients.
    \item Follow Users --- users can select another user to ``follow'' which will synchronize the user's camera with the web or HoloLens user's view.
\end{itemize}

\subsubsection{Renderer}
\label{sec:web-renderer}
The central technology behind the session viewer is the layered 3D model renderer built on top of Three.js. Unlike the HoloLens, the web renderer was designed to visualize extremely large models with less of a focus on real-time rendering performance. Driving this decision was the need to render models in their original complexity (5-50 million polygons) on the web so that users could perform reductions to optimize the models for HoloLens viewing (see Section~\ref{sec:model-reduction}).
The ProtoSpace 3D renderer has been open sourced~\cite{psCADrenderer}. Here we describe key aspects of the renderer:
\begin{itemize}
    \item Iterative Node Drawing --- the renderer operates on a triangle budget per render call. When the model needs to be drawn, the renderer flattens the model tree into a list, sorts the nodes by triangle count, and draws the maximum number of nodes within the triangle limit over successive frames until all nodes have been drawn. This approach aims to provide a smooth user interaction with the 3D model that will eventually be fully drawn over several frames. When a user changes the camera position or the model geometry changes,  model rendering will redraw from the beginning.
    \item Compositing Layers --- the renderer draws to several independent layers and then composites these layers in order to only update certain layers when necessary. These layers include the model layer, a node highlight layer, an annotation layer where the points of interest, HoloLens avatars, and AprilTags are drawn, and a floor layer. All layers besides the model layer are drawn each frame as these interactions are primarily driven by real-time user behavior.
    \item View Cube --- the renderer draws a 3D view cube for ease of selecting camera orientation. This view cube has been open sourced~\cite{viewcube}.
\end{itemize}

\section{Case Studies}

In this section, we highlight several case studies with quotes from end users to showcase the real benefit gained by ProtoSpace users.

\subsection{Mars 2020}
\begin{figure}
    \centering
    \includegraphics[width=1\linewidth]{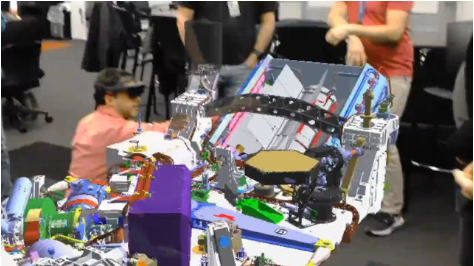}
    \caption{Mars 2020 mission~\cite{farley2020mars} discussion session in ProtoSpace; shown here is the top portion of the Mars Perseverance rover in the stowed position.}
    \label{fig:m20-opslab}
\end{figure}

The Mars 2020 mission~\cite{farley2020mars} used ProtoSpace for analyzing potential issues during the entry decent landing phase of the mission. See Figure~\ref{fig:m20-opslab}.
\begin{displayquote}
``\textit{We were looking for potential areas on the rover that could snag with our cabling as they're retracted from the rover top deck. Though one can look at a flat CAD drawing in a meeting, ProtoSpace allowed a large group to interact with the CAD mock-up of the top deck simultaneously and from different angles. It also gave a better sense of the physicality of the problem. I thought it was a perfect use case for the system.}'' --- Entry Decent Landing (EDL) Systems Engineer, Mars 2020
\end{displayquote}

\subsection{Europa Clipper}
\begin{figure}
    \centering
    \includegraphics[width=1\linewidth]{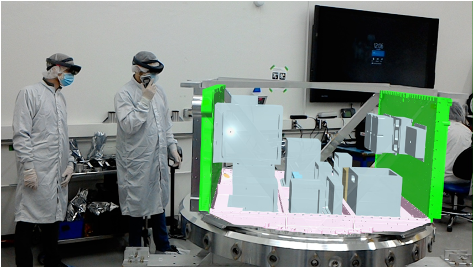}
    \caption{Europa Clipper~\cite{pappalardo2024science}  session in ProtoSpace focused on the avionics vault.}
    \label{fig:clipper-vault}
\end{figure}

\begin{figure}
    \centering
    \includegraphics[width=1\linewidth]{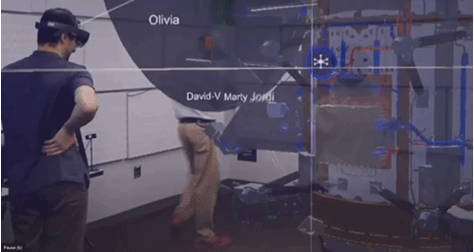}
    \caption{A non-collocated session between engineers at the Applied Physics Lab and JPL, working on the Europa Clipper mission~\cite{pappalardo2024science}. Remote users' names and gaze directions were shown to the local participants in augmented reality within ProtoSpace, while they all viewed the same CAD model for a collaborative engineering discussion.}
    \label{fig:apl-jpl}
\end{figure}

The Europa Clipper mission~\cite{pappalardo2024science} used ProtoSpace for early design, especially for the avionics vault as well as for cross-country remote collaboration between JPL and the Applied Physics Lab. See Figures~\ref{fig:clipper-vault},~\ref{fig:apl-jpl}, and~\ref{fig:clipper-pdr}.

\begin{displayquote}
``\textit{The Avionics Vault of the Europa Clipper Spacecraft is essentially a 3-D puzzle box. Protospace gives us the chance to swiftly solve this 3-D puzzle and to optimize the design!}'' --- Mechanical Engineer, Europa Clipper 
\end{displayquote}

\subsection{NISAR}
\label{sec:nisar}
The NISAR mission~\cite{kellogg2020nasa} used ProtoSpace extensively for all three of the main use cases that ProtoSpace supports---spacecraft design, integration \& test, and presentations (see Section~\ref{sec:three-main-use-cases}).

\begin{displayquote}
\textit{``Aerospace components are designed and built to be lightweight, volume constrained, efficient structures with streamlined load paths. However, that comes at a cost and all too often suffers when it comes to component assembly. Before any components have even begun fabrication, ProtoSpace can be used to visualize and predict the assembly process, preventing complications with hardware while still in the design phase. Pictures say 1000 words, videos say even more. What do you think holograms overlaid on the physical world say?''} --- Mechanical Integration Engineer, NISAR
\vspace{0.5em}
\end{displayquote}

\begin{displayquote}
\textit{``Before a big lift, we put the headsets on and we can move something into place by projecting it and showing people what the working area will be like so we can understand if there will be any issues or close clearances. We’ve had some really close clearance lifts and we bide down risk by understanding the intricacies before we have two tons of hardware on a crane hook.''} --- Mechanical I\&T Lead, NISAR
\end{displayquote}

\begin{figure}
    \centering
    \includegraphics[width=1\linewidth]{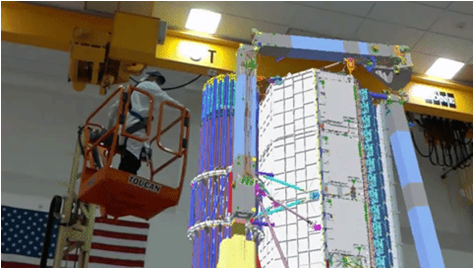}
    \caption{Here, an engineer wore the HoloLens headset to visualize a yet-to-be-built NISAR spacecraft~\cite{kellogg2020nasa}, while on a lift platform to determine if it would work for a future procedure.}
    \label{fig:nisar-cleanroom}
\end{figure}

\subsection{CGI}
\begin{figure}
    \centering
    \includegraphics[width=1\linewidth]{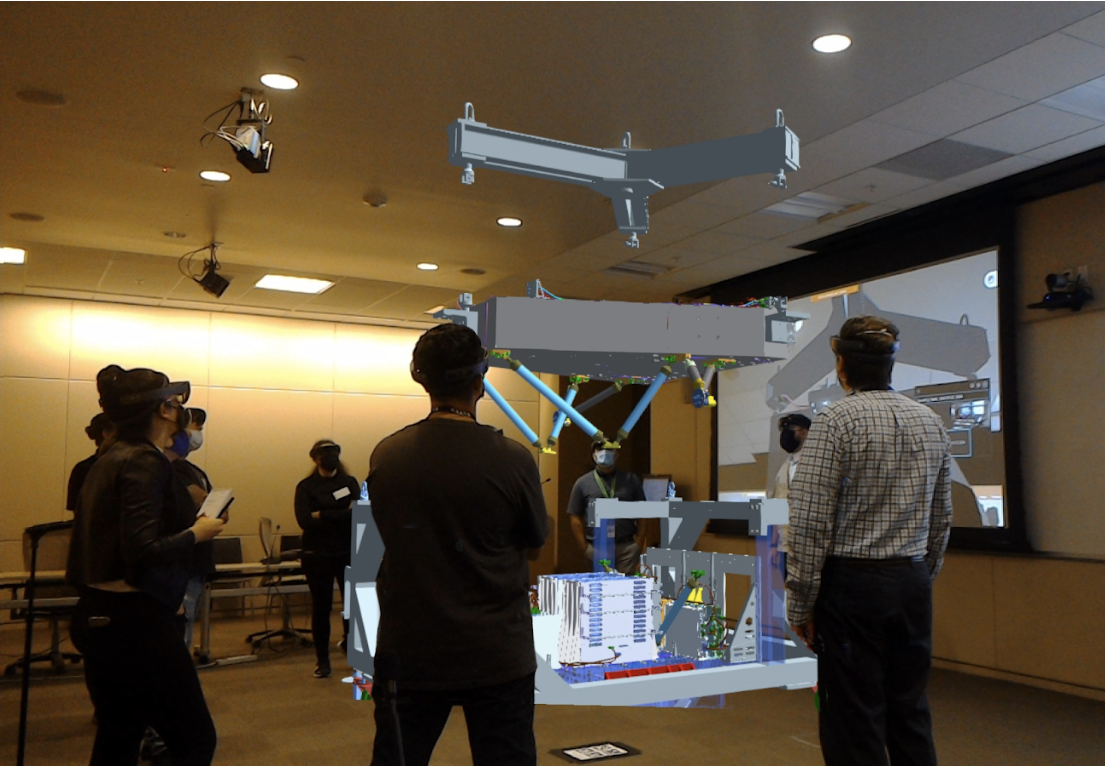}
    \caption{Coronagraph Instrument (CGI) Instrument I\&T Discussion~\cite{bedrosian2025roman}, part of the Nancy Grace Roman Space Telescope (NGRST) mission.}
    \label{fig:cgi-int-planning}
\end{figure}

Bedrosian et al.~\cite{bedrosian2025roman} describe in detail how they used ProtoSpace for their instrument I\&T for the Roman Coronagraph Instrument (CGI)~\cite{bailey2023nancy}. See Figures~\ref{fig:roman-pdr} and~\ref{fig:cgi-int-planning}.

\begin{displayquote}
\textit{``The session was an immense success, and the I\&T Team was more than satisfied with the capabilities of the technology... The efficiency of performing storyboards in ProtoSpace relative to NX is on the scale of 25\% of the time.''} --- I\&T Mechanical Team, CGI 
\end{displayquote}

\subsection{SPHEREx}
\label{sec:spherex}
The SPHEREx mission~\cite{crill2020spherex} used ProtoSpace for enhanced communication during their Site Visit Proposal.

\begin{displayquote}
\textit{``ProtoSpace gave us a unique and high-tech platform to share the elegant simplicity of our design and engage the evaluation team one-on-one during our site visit.''} --- Project Systems Engineer, SPHEREx
\end{displayquote}

\subsection{SRL}
ProtoSpace is currently being used in early designs for the Sample Retrieval Lander (SRL) portion of the planned Mars Sample Return program~\cite{muirhead2020mars}; it was also used in the Sample Recovery Helicopter (SRH) concept~\cite{mier2023sample} that is no longer in the program baseline.

\begin{displayquote}
\textit{``While routing cables on the access door on the cone we needed to understand the best way to route the cables to a bracket, what the hand access would be like, if we should have 1 bracket or 2 and so on. We talked about making a mockup but instead we got a few headsets, loaded a session and were able to put our own hands in to see for ourselves.''} --- Cable Harness Engineer, SRL
\vspace{0.5em}
\end{displayquote}

\begin{displayquote}
\textit{``I did the routing for SRH and that was difficult from the beginning. A lot of thick cables that had to route through a tiny cramped space with multiple moving parts that we had to connect to but also stay clear of, and on top of all that, the entire box had to hinge 180°. Once we had the basic concept of the routing we needed to ensure that there wouldn’t be any problems with the rotation. We made a mockup and I was able to overlay the model on the headset to the actual mockup so we could see it in front of us and identify any possible snags.''} --- Cable Harness Engineer, SRL
\end{displayquote}

\label{sec:one-offs}
\section{One-Offs}
This section describes several offshoots of ProtoSpace that do not fall within the main three uses of ProtoSpace as described in Section~\ref{sec:three-main-use-cases}.

\subsection{Team Xc Virtual Whiteboard}
\label{sec:teamxc}
\begin{figure}
    \centering
    \includegraphics[width=1\linewidth]{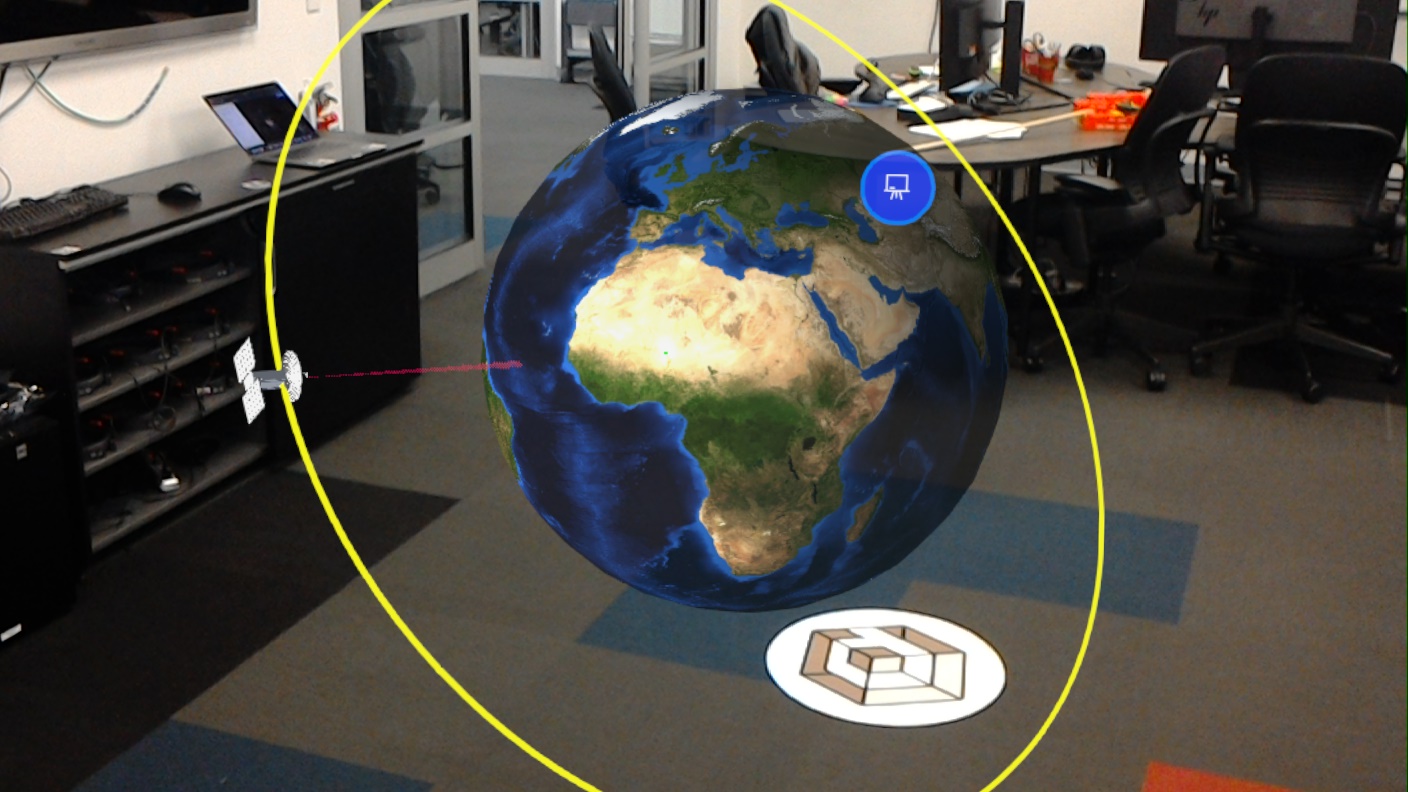}
    \caption{Virtual Whiteboard showing the RainCube concept~\cite{peral2019raincube} in orbit above Earth.}
    \label{fig:virtualwhiteboard}
\end{figure}

The Virtual Whiteboard is a session configuration option of ProtoSpace intended to provide a first-pass look at whether early mission design configurations (i.e. spacecraft configuration, orbit trajectory) will achieve the intended mission goals. Originally designed for JPL's rapid, concurrent mission concept team for small satellites, TeamXc, the Virtual Whiteboard is developed to provide physically accurate and to-scale representations and is available on both the Web Portal and the HoloLens. The Virtual Whiteboard enables users to load a planetary body, a CAD model, the Sun, a field of view frustum, and a default orbit configuration (e.g. Low Earth Orbit) to match the mission design concept. The Virtual Whiteboard Configuration is highly customizable; the planetary body locations, orbit configuration, satellite orientation, and the field of view parameters may be edited by user command. See Figure \ref{fig:virtualwhiteboard} to see the RainCube concept~\cite{peral2019raincube} in orbit above Earth using the Virtual Whiteboard.

\subsection{QMDT}
\begin{figure}
    \centering
    \includegraphics[width=1\linewidth]{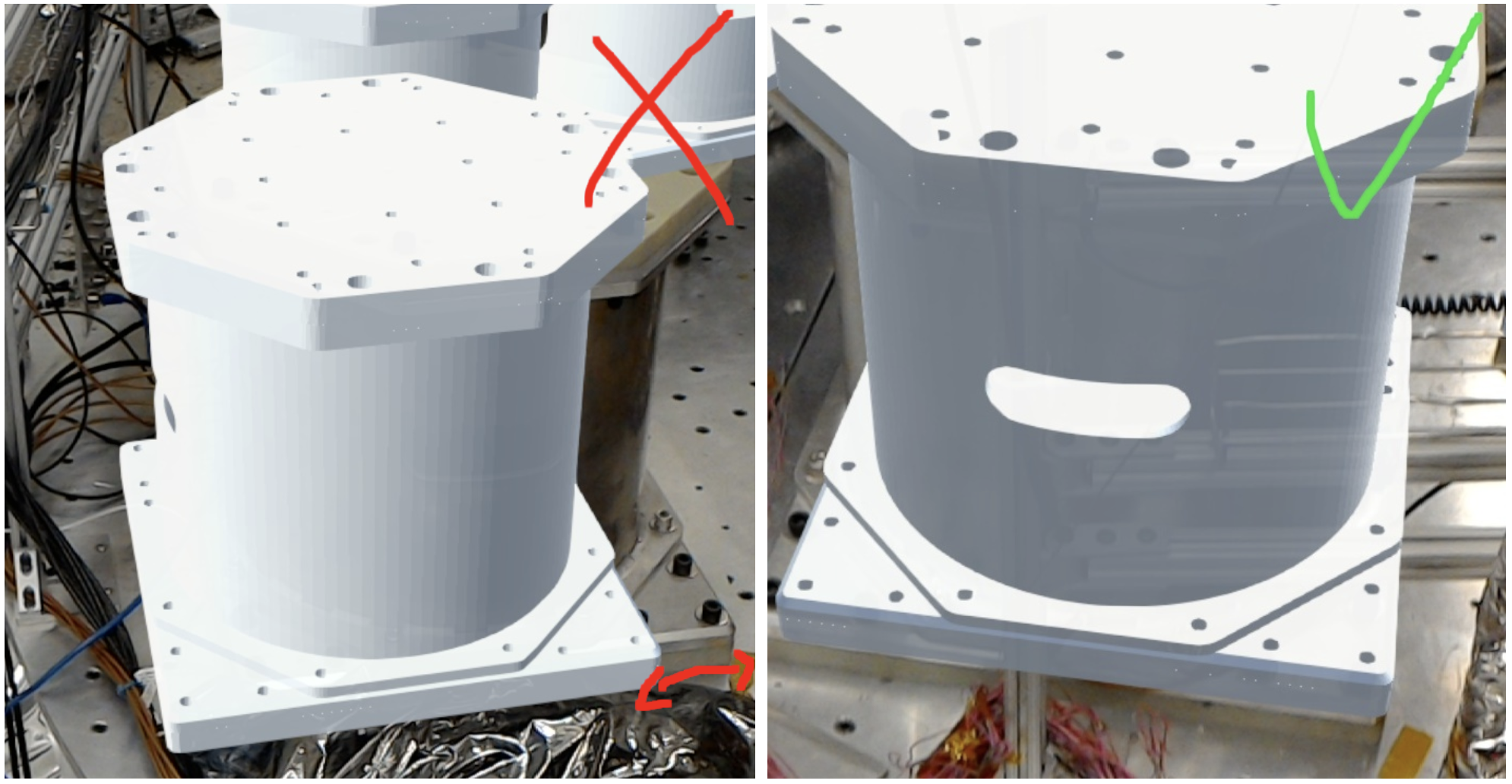}
    \caption{QMDT~\cite{megivern2022simulating} chamber configuration validation; left: misalignment caught using ProtoSpace; right: good configuration.}
    \label{fig:qmdt}
\end{figure}

\begin{figure}
    \centering
    \includegraphics[width=1\linewidth]{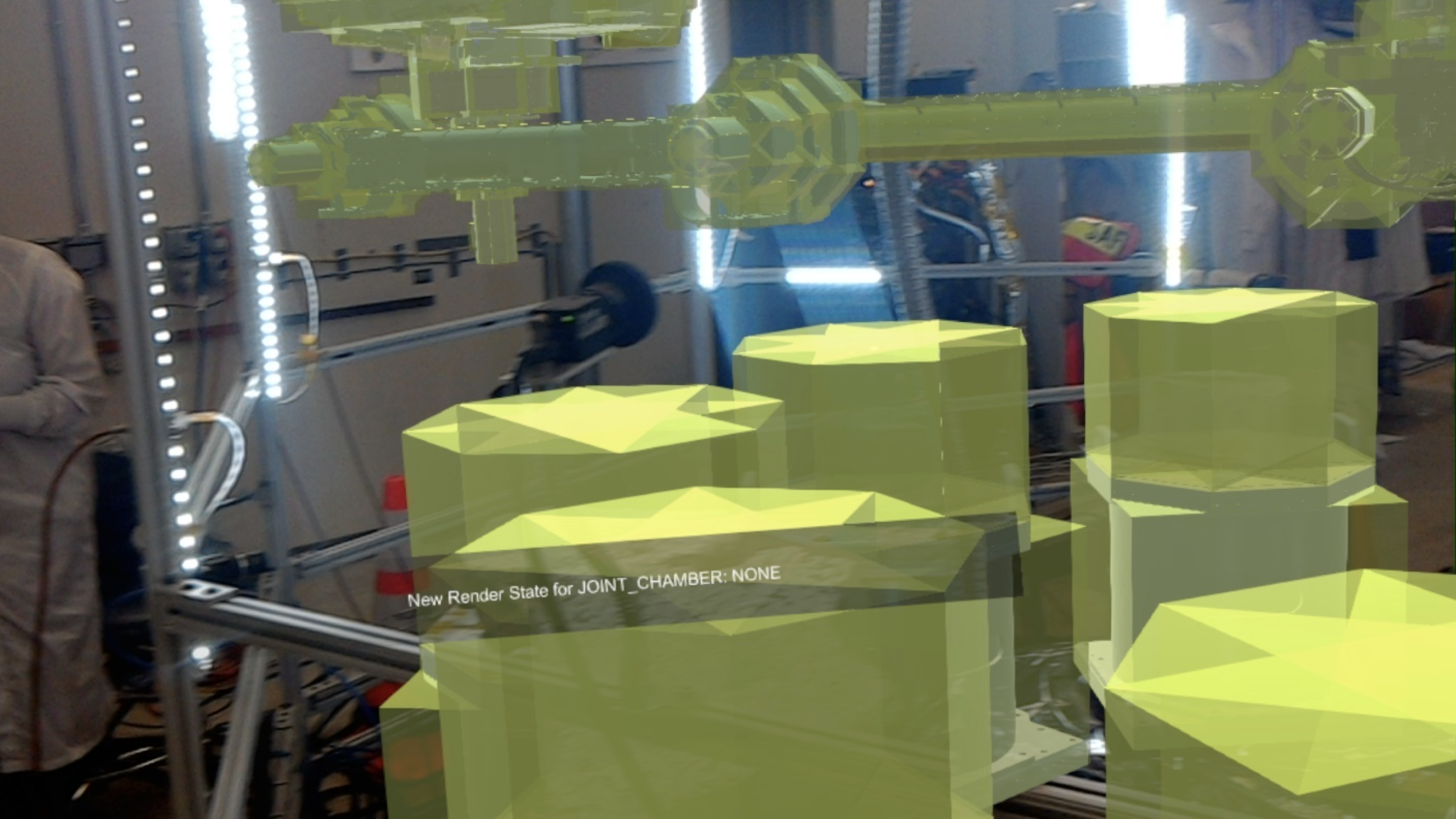}
    \caption{QMDT~\cite{megivern2022simulating} collision volumes were visualized for ensuring proper chamber configuration.}
    \label{fig:collision-volumes}
\end{figure}

Portions of ProtoSpace were used for the Quality Model Dirty Testing (QMDT) of the Mars 2020 mission~\cite{megivern2022simulating}.
In particular, the testing configurations for drilling rocks were shown in the HoloLens for the testing technician to confirm proper environment testing setup.
In one instance, engineers caught an incorrect configuration (see Figure~\ref{fig:qmdt}) that, had it gone unnoticed, could have caused damage to the testing environment, which would have ultimately caused delays in the mission.
Collision volumes were also visualized, as shown in Figure~\ref{fig:collision-volumes}, helping engineers and technicians confirm a proper chamber configuration.

\subsection{Procedures}
\label{sec:procedures}

In partnership with the Johnson Space Center (JSC), JPL participated in the Mixed Reality Crew Assistance (MRCA) project~\cite{mrca}.
This led to two investigations, detailed in the next two sections, related to using ProtoSpace for procedure assistance for the Cold Atom Lab (CAL) and for the NASA Extreme Environment Mission Operations (NEEMO) mission.

\subsubsection{CAL experiment}
\begin{figure}
    \centering
    \includegraphics[width=1\linewidth]{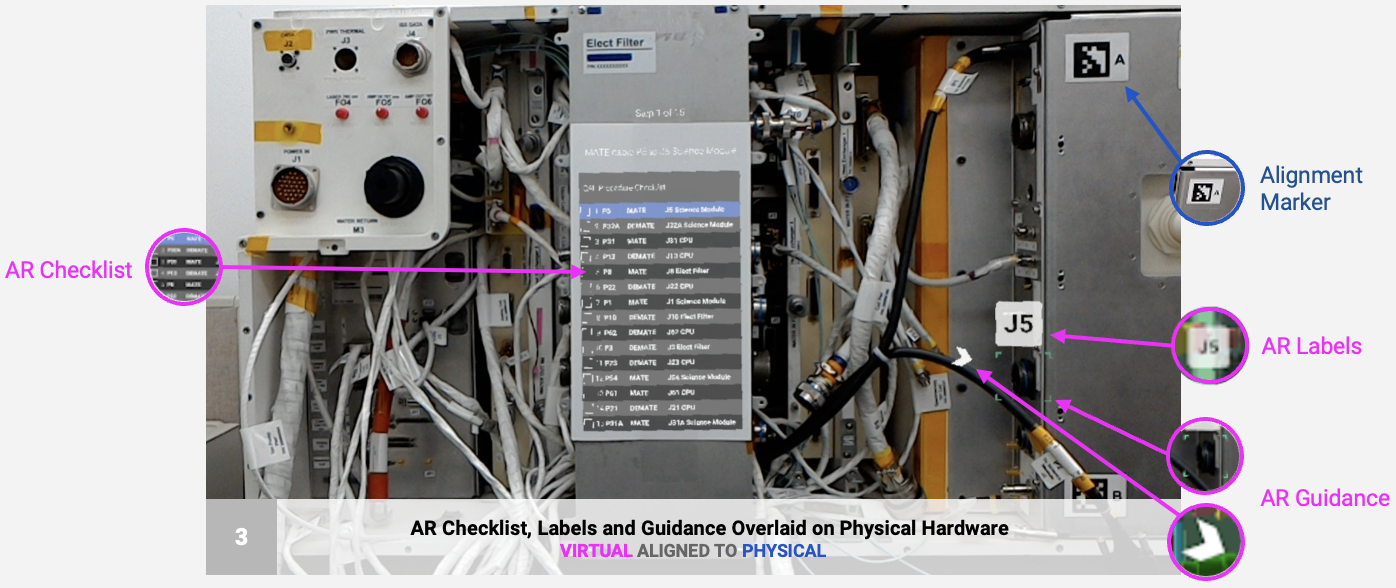}
    \caption{The procedure interface for the CAL experiment.}
    \label{fig:cal-procedure}
\end{figure}

The Cold Atom Lab has allowed scientists to conduct quantum physics experiment in the micro-gravity environment onboard the ISS~\cite{aveline2020observation}.
To develop and test our AR procedure interface, we used an engineering mockup of CAL as a testing ground.
Figure~\ref{fig:cal-procedure} shows the augmented reality interface used to guide someone to perform maintenance.
A user experiment showed faster task completion times and lower mental workload than a traditional paper procedure method~\cite{braly2019augmented}.

\subsubsection{NEEMO experiment}
ProtoSpace was used as part of the NEEMO undersea, ISS analog mission~\cite{nuernberger2020under}.
Four crew members successfully used our step-by-step AR procedure interface

\subsection{Animated Instructions}
Over the years, there have been a variety of situations where customers needed simple 3D animated instructions or views of CAD models.
We used the underlying ProtoSpace framework to develop each of these.

\subsubsection{VITAL Ventilator}
\label{sec:vital}
\begin{figure}
    \centering
    \includegraphics[width=1\linewidth]{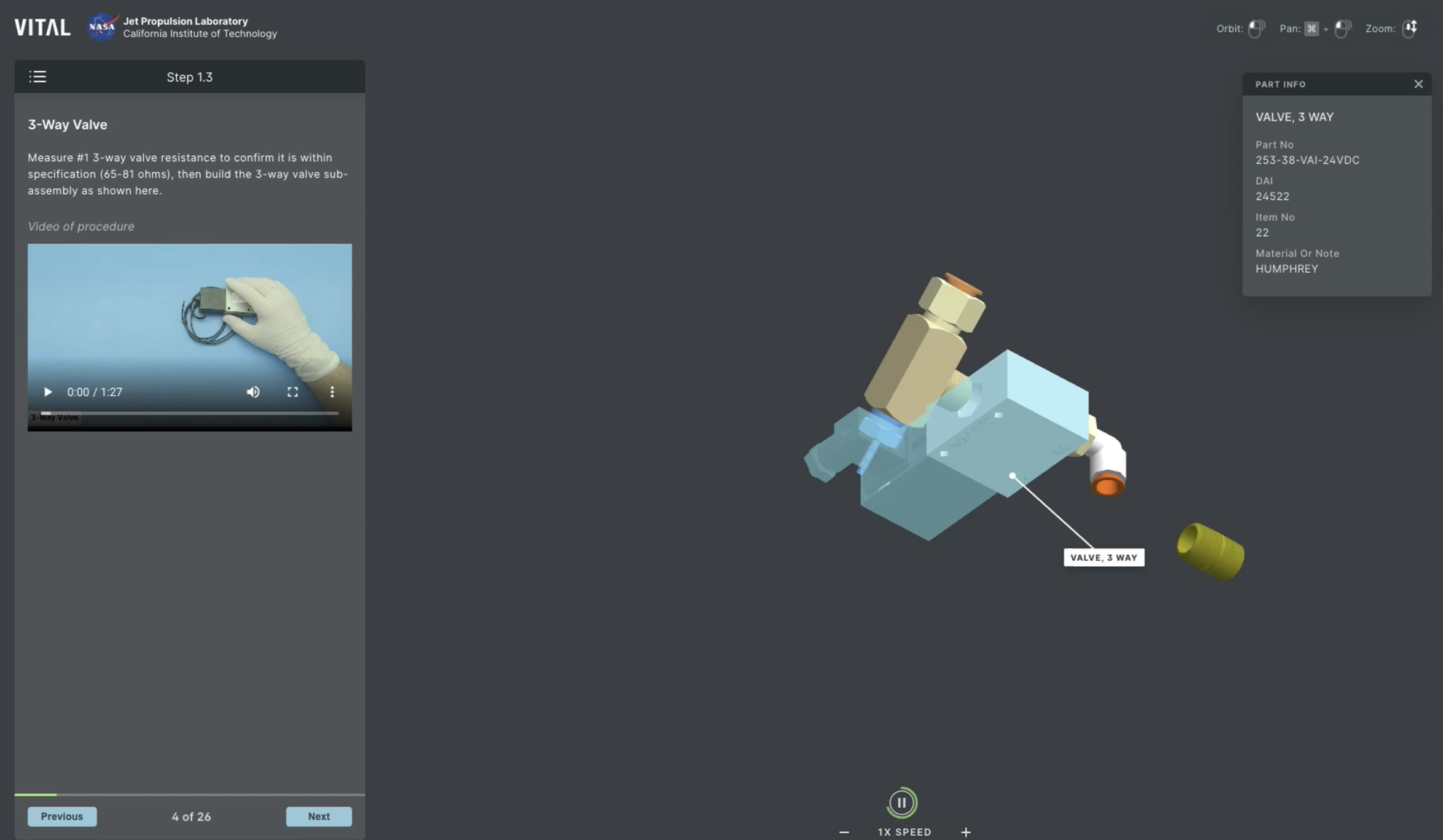}
    \caption{ProtoSpace was used to create animated assembly instructions for the VITAL Ventilator~\cite{hill2020vital}.}
    \label{fig:vital}
\end{figure}

The VITAL (Ventilator Intervention Technology Accessible Locally) project utilized ProtoSpace to create interactive assembly and operations instructions for a high-pressure ventilator designed to treat COVID-19 patients~\cite{hill2020vital}. Approved by the FDA under Emergency Use Authorization, the device was tailored to free up traditional ventilators for patients with the most severe symptoms. Using the ProtoSpace Web Viewer, the team created virtual instructions in 3.5 weeks, allowing licensees worldwide to confirm assembly steps and enabling respiratory therapists and nurses to access operation information from their mobile devices. This application of ProtoSpace's framework established a foundation for subsequent interactive visualization tools. See Figure~\ref{fig:vital}.

\subsubsection{EELS}
\begin{figure}
    \centering
    \includegraphics[width=1\linewidth]{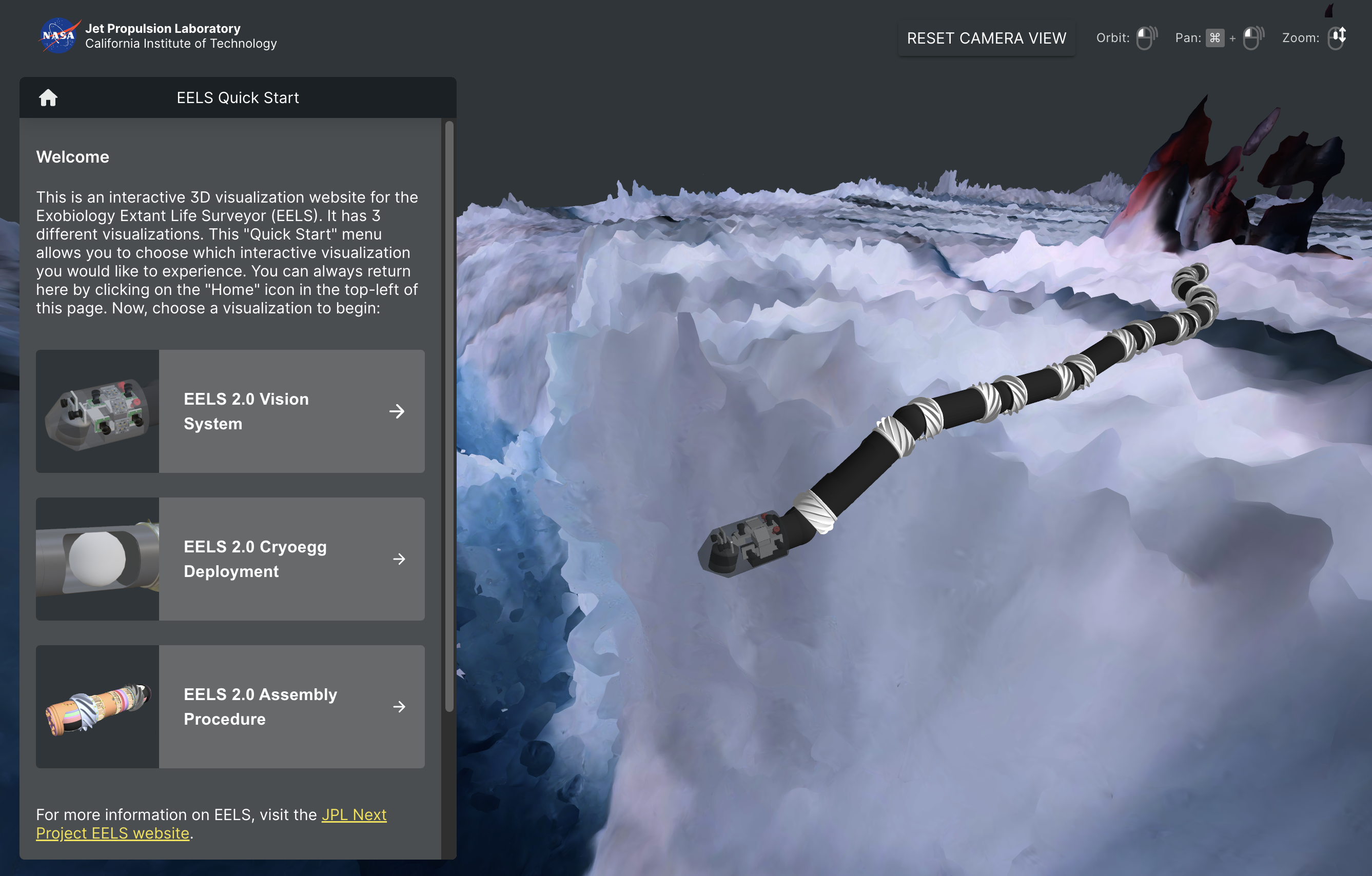}
    \caption{The EELS mission~\cite{vaquero2024eels} used ProtoSpace for showcasing the EELS 2.0 Vision System, the Cyroegg Deployment, and the assembly of the robot.}
    \label{fig:eels}
\end{figure}

The EELS mission~\cite{vaquero2024eels} used ProtoSpace to showcase the assembly procedure for the EELS 2.0 robot, as well as how the EELS mission would deploy the robot on a glacier field trip. Figure~\ref{fig:eels} shows how the EELS robot could potentially be anchored nearby a crevasse.

\subsubsection{CAL GIFs}
\label{sec:cal-gifs}
\begin{figure}
    \centering
    \includegraphics[width=.7\linewidth]{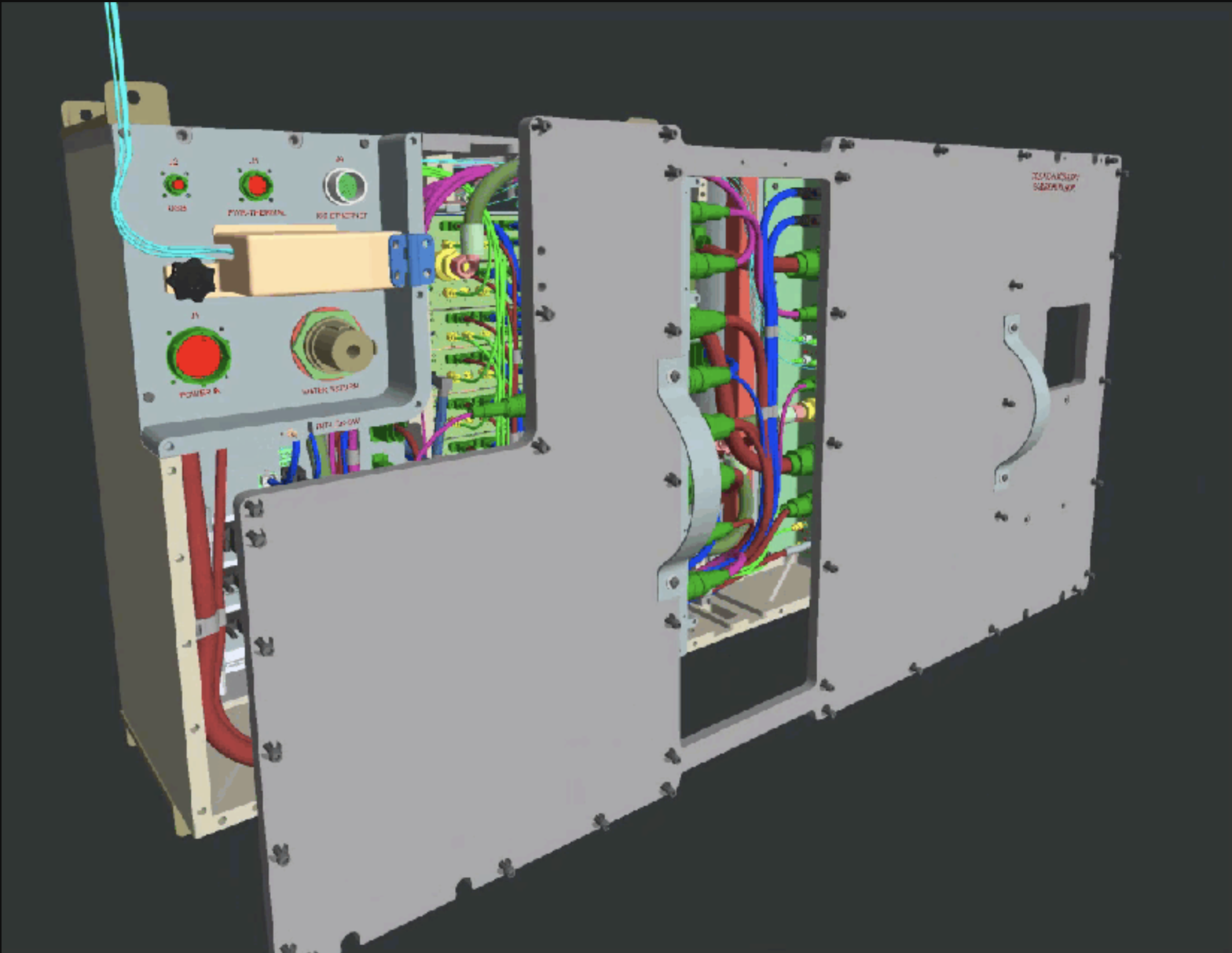}
    \caption{One frame of an animated GIF produced by ProtoSpace and used in a maintenance procedure for the Cold Atom Lab~\cite{aveline2020observation,kellogg2023augmented}.}
    \label{fig:cal-gif-frame}
\end{figure}

The Cold Atom Lab~\cite{aveline2020observation,kellogg2023augmented} used ProtoSpace for producing animated GIFs for the procedural instructions given to an ISS astronaut for a maintenance procedure. This was the first time that animated GIFs were used in ISS payload procedures. Figure~\ref{fig:cal-gif-frame} shows an illustration.

\subsubsection{Power to Explore Experience}
\begin{figure}
    \centering
    \includegraphics[width=1\linewidth]{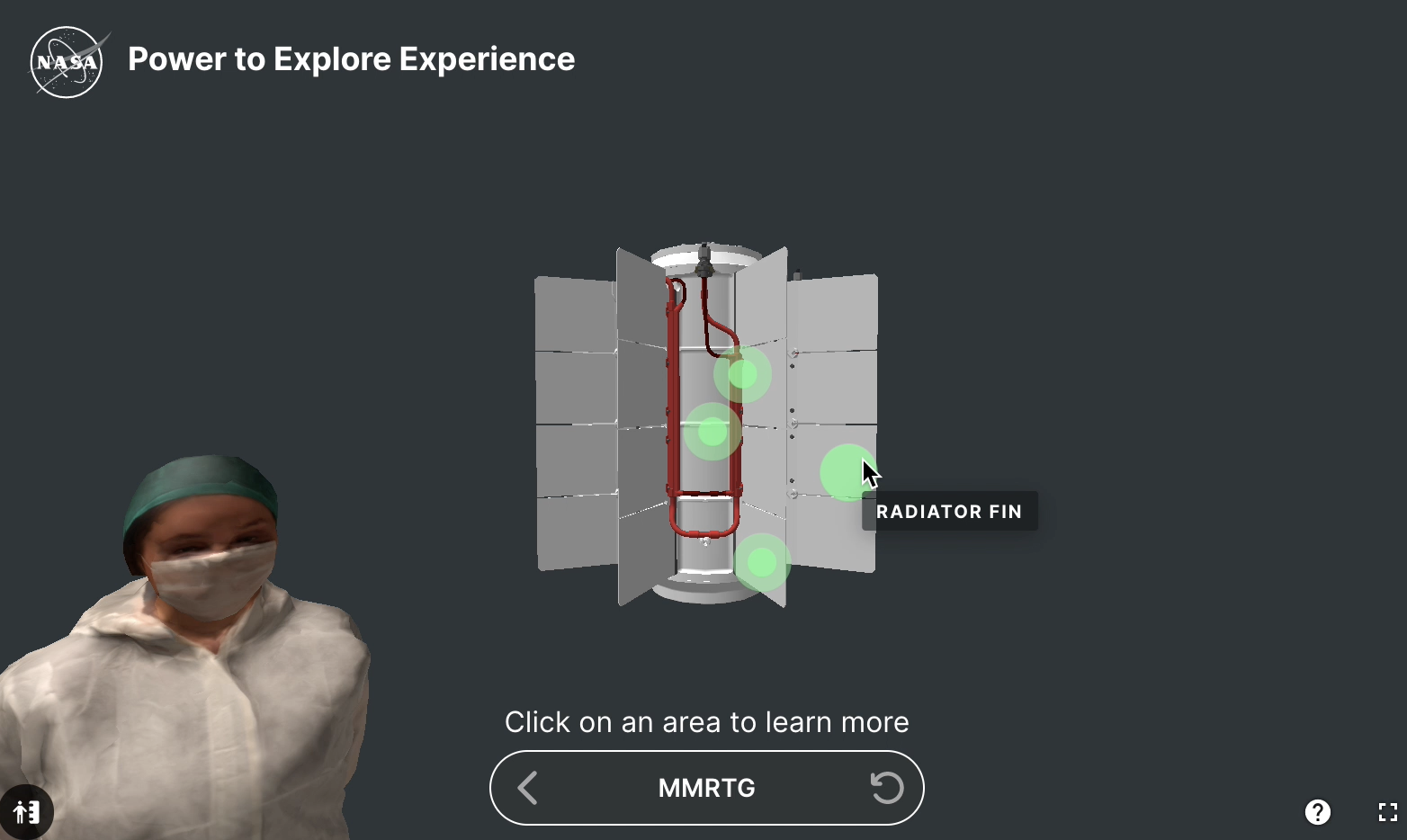}
    \caption{Power to Explore Experience showing the Multi-Mission Radioisotope Thermoelectric Generator.}
    \label{fig:PowerToExplore}
\end{figure}

Built in collaboration with NASA Glenn Research Center, the Power to Explore Experience is a web-based visualization tool intended to provide details about how Mars 2020 is powered using the Multi-Mission Radioisotope Thermoelectric Generator (MMRTG). The Power to Explore Experience was built using the ProtoSpace messaging system, 3D Web Viewer, and step-by-step assembly framework developed for the VITAL Ventilator Assembly project. Users are provided a user interface to transition between different view configurations; starting with the Mars 2020 Rover, the user may transition to viewing solely the MMRTG with clickable points of interest that orient the view for a better look and provides text call-outs to learn about the components of the MMRTG. A toggleable human CAD model is provided for scale. See Figure \ref{fig:PowerToExplore} for a look at the Power to Explore Experience.
\cite{rps3dviewer}

\section{Lessons Learned}

\subsection{AR over VR}
\begin{figure}
    \centering
    \includegraphics[width=1\linewidth]{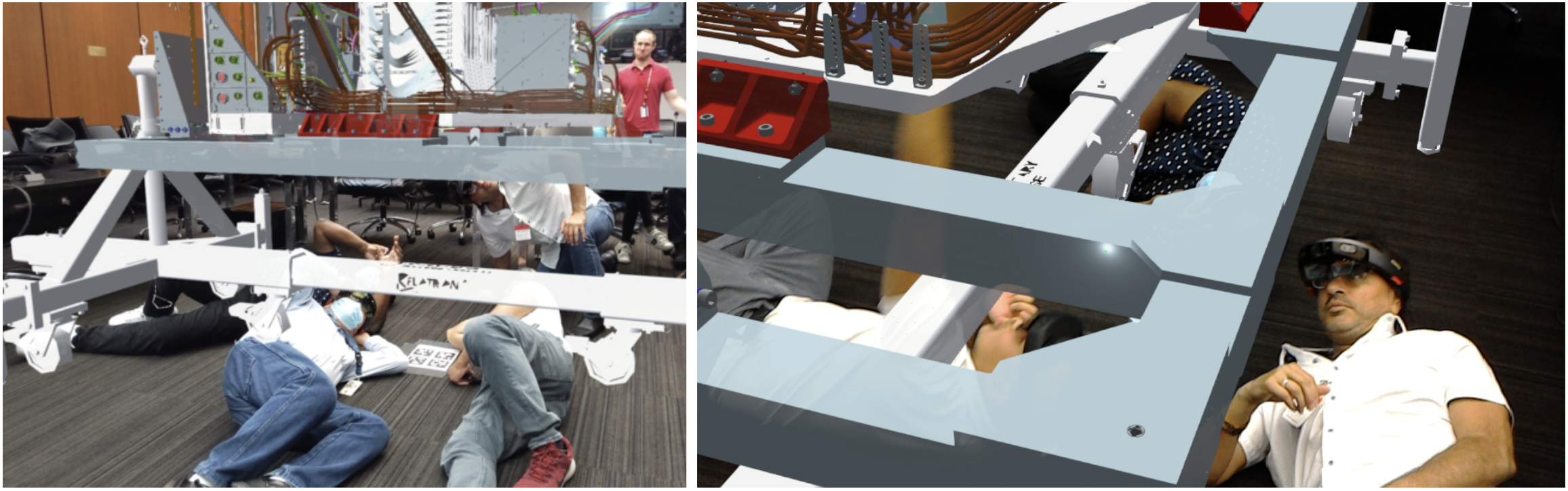}
    \caption{Engineers planning integration and test on spacecraft hardware. Notice how using untethered optical-see-through AR displays allowed engineers to quickly and safely assume a position underneath the virtual CAD model.}
    \label{fig:ar-over-vr}
\end{figure}

Some immersive CAD visualization tools focus on fully virtual (VR) experiences; this is mainly due to the necessities of the use case, such as needing to be immersed into an architecture design.
However, with ProtoSpace's main use cases (see Section~\ref{sec:three-main-use-cases}), we observed that augmented reality was sufficient and more desired than VR.
There are several reasons why engineers preferred the AR approach.
First, it is extremely difficult, if not impossible, to duplicate the physical context in VR. 
With augmented reality, users can see the physical cleanroom in which spacecraft is being assembled; users can see their colleagues, their physical hardware, and tools; see Figure~\ref{fig:ar-over-vr}.
With optical-see-through augmented reality, there is also less risk of motion sickness due to users' being able to directly see the physical world.
For the same reasons, AR is likely much safer than VR as well.
Finally, we also observed that users new to XR tend to enjoy using AR headsets over VR headsets, again for the same reasons---users can feel safer when seeing the real world.

\subsection{Business Case}
\footnote{The cost information contained in this document is of a budgetary and planning nature and is intended for informational purposes only. It does not constitute a commitment on the part of JPL and/or Caltech.}We conducted several interviews with JPL Mechanical Engineers who used ProtoSpace for NISAR~\cite{kellogg2020nasa}, Europa Clipper~\cite{pappalardo2024science}, and CGI~\cite{bedrosian2025roman}.
From this, we estimate an average of 12.75 days of savings in work per year per engineer for doing small configuration changes as a result of utilizing ProtoSpace.
Larger design changes could reap much larger benefits, especially when design issues are fixed, avoiding re-work later on (e.g., Europa Clipper saved an estimated \$1-2M based on making an early design change after using ProtoSpace).

Finally, we note that ProtoSpace was the JPL Software-of-the-Year Award (SOYA) Winner in 2021. It also received an Honorable Mention in the NASA SOYA competition in the same year.

\section{Future Outlook}
While there are several commercial products now available that readily showcase CAD models in augmented or virtual reality, ProtoSpace is still in use at JPL mainly due to its having organically evolved to fit the specific needs of the various engineers working on designing and building spacecraft at JPL.
Currently, ProtoSpace runs on HoloLens 1 and HoloLens 2 headsets, and some features have been ported to the Meta Quest platform~\cite{berndt2023universe}; future work may involve porting ProtoSpace to newer headsets.
Another area of possible future work is tighter integration with the the original CAD systems, as noted in Section~\ref{sec:related-work-xr-cad-viewing}.
In terms of augmented reality technical needs, while many AR platforms now support ``object tracking'' alignment capabilities, there is less support for alignment to partially built spacecraft where CAD models don't accurately represent the dynamically changing nature of the physical hardware.

\section{Conclusion}
In conclusion, we presented a review of the ProtoSpace system that has been used at JPL for the past 10 years.
When ProtoSpace was first created, nothing like it existed.
Today, there are several similar commercial products, yet ProtoSpace still brings strong value to the engineers at JPL and is still in use today, a testament to both the software engineering and user interface design of ProtoSpace as well as the longevity of the HoloLens hardware itself.

The main technical value-add is providing the ability to see 3D CAD stereoscopically and collaboratively in untethered, optical-see-through augmented reality.
This ultimately changes the style of communication---while engineers may have previously all viewed a presentation on a projector slide screen, all sitting around a table in a conference room, engineers can now stand around a virtual CAD model and have a potentially more engaged discussion, physically gesturing to one another with respect to the virtual model.
Another value-add ProtoSpace provides is a simple model reduction interface (e.g., in some cases, people just used ProtoSpace for model reduction only, no AR at all).

Finally, numerous times we have seen end-user engineers fairly quickly become ``power users'' of ProtoSpace, using the product all by themselves without any hands-on assistance from the team that made the ProtoSpace system.
All of this is a testament to the versatility of both the system, as well as the people who designed, developed, and deployed ProtoSpace throughout JPL.


%



\section*{Acknowledgments}
Over forty individuals have contributed directly to the development of ProtoSpace throughout the past ten years and numerous individuals and projects have advocated for its advancement.
We particularly acknowledge the management who were early champions for the initial concept as well as the original team who designed and built the core algorithms and user interface that have allowed ProtoSpace to last such a long time.
We also acknowledge external partnerships that have been crucial to the success of ProtoSpace, including the Microsoft HoloLens team for their continued support and advice during early tool development; our colleagues at JSC, especially in regard to immersive procedures~\cite{mrca}; and our colleagues at the Applied Physics Lab who have utilized their own instance of ProtoSpace and provided much valuable feedback~\cite{hanhe2018risk}.
This work was carried out at the Jet Propulsion Laboratory, California Institute of Technology, under a contract with the National Aeronautics and Space Administration (80NM0018D0004).

\ifCLASSOPTIONcaptionsoff
  \newpage
\fi



%

\bibliography{references}{}
\bibliographystyle{plain}

%








\end{document}